\documentclass[a4paper,11pt]{article}
\pdfoutput=1
\usepackage{amsmath,amssymb,mathtools}
\usepackage{placeins} 

\usepackage[table]{xcolor}
\usepackage{tcolorbox} 
\definecolor{mygray}{gray}{0.2}
\renewcommand*{\arraystretch}{2.0}

\allowdisplaybreaks

\usepackage{jheppub}
\usepackage{bm,amssymb,slashed,graphicx,multirow,mathtools,xspace,array}  
\usepackage{float} 
\restylefloat{table}                  
\usepackage{subfigure}
 \usepackage{slashed}
 \usepackage{acronym}
 \usepackage{amsfonts}
 \usepackage{tensor}
\definecolor{violet}{rgb}{0.94, 0.2, 0.8}%lightcoral
\definecolor{lightblue}{rgb}{0.29, 0.28, 0.93} %cornflowerblue
\definecolor{asparagus}{rgb}{0.13, 0.6, 0.12}

\newcommand{\com}[1]{{#1}}

\newcolumntype{C}{>{$}c<{$}} % math-mode version of "c" column type    
\newcolumntype{L}{>{$}l<{$}} % math-mode version of "l" column type    
\newcolumntype{R}{>{$}r<{$}} % math-mode version of "r" column type   

\newcommand{\szerot}{\tilde{s}_0}
\newcommand{\tzerot}{\tilde{t}_0}
\newcommand{\Sredhat}{ \hat{\sigma}^{(a)}_0}

\newcommand{\Sred}{ {\sigma}^{(a)}_0}
\newcommand{\Sreda}[1]{ {\sigma}^{(#1)}_0}
\newcommand{\bsup}[1]{ \bar{\de}^{(a)}_s(#1)}  
\newcommand{\btup}[1]{ \bar{\de}^{(a)}_t(#1)}

\acrodef{PDG}[PDG]{Particle Data Group}
\acrodef{OPE}[OPE]{Operator Product Expansion}
\acrodef{FCNC}[FCNC]{flavour-changing neutral current}
\acrodef{RHC}[RHC]{right-handed currents}
\acrodef{SM}[SM]{Standard Model}
\acrodef{NP}[NP]{New Physics}
\acrodef{MFV}[MFV]{Minimal Flavour Violation}
\acrodef{SD}[SD]{short-distance}
\acrodef{LD}[LD]{long-distance}
\acrodef{DA}[DA]{distribution amplitude}

\newcommand{\den}{S}

 \newcommand{\CondG}{{\left \langle  G^2 \right \rangle}}

\newcommand{\CondQQ}[1]{{ \left\langle \bar{#1} #1 \right \rangle}}

\newcommand{\CondqGq}[1]{{ \left \langle \bar{ #1 }  G #1 \right \rangle }}

\newcommand{\type}{\Gamma}

\newcommand{\VEV}[1]{\left\langle #1 \right\rangle} 
\newcommand{\vev}[1]{\langle #1 \rangle}

\newcommand{\para}{\parallel}
\newcommand{\matel}[3]{\langle #1|#2|#3\rangle}
\newcommand{\al}{\alpha}

\newcommand{\be}{\beta}
\newcommand{\ga}{\gamma}
\newcommand{\de}{\delta}

\newcommand{\eps}{\epsilon}

\newcommand{\Ga}{\Gamma}

\newcommand{\Cdot}{\!\cdot\!}
\newcommand{\mi}{\!-\!}
\newcommand{\pl}{\!+\!}

\newcommand{\GeV}{\,\mbox{GeV}}
\newcommand{\MeV}{\,\mbox{MeV}}
\newcommand{\keV}{\,\mbox{keV}}

\newcommand{\UU}{U}

\newcommand{\Rea}{\textrm{Re}}
\newcommand{\Ima}{\textrm{Im}}
\newcommand{\disc}{\textrm{disc}}
\newcommand{\tk}{\tilde{k}}

\newcommand{\gonepl}{g_{BB_1\gamma}}
\newcommand{\gonemi}{g_{BB^*\gamma}}

\newcommand{\Pperp}[1]{P^\perp_{#1}}
\newcommand{\Ppara}[1]{P^\parallel_{#1}}
\newcommand{\Pperppara}[1]{P_{\perp,\parallel}^{#1}}

\newcommand{\FV}{V_\perp}
\newcommand{\FA}{V_\parallel}

\newcommand{\FTV}{T_\perp}

\newcommand{\FTA}{T_\parallel}

\newcommand{\mDz}{m_{D^0}}
\newcommand{\mDd}{m_{D^+}}
\newcommand{\mDs}{m_{D_s}}
\newcommand{\mBz}{m_{B^0}}
\newcommand{\mBu}{m_{B^+}}
\newcommand{\mBs}{m_{B_s}}

\newcommand{\mDsz}{m_{D^{*0}}}
\newcommand{\mDsd}{m_{D^{*+}}}
\newcommand{\mDss}{m_{D^*_s}}
\newcommand{\mBsz}{m_{B^{*0}}}
\newcommand{\mBsu}{m_{B^{*+}}}
\newcommand{\mBss}{m_{B^*_s}}
\newcommand{\mB}{m_B}

\newcommand{\Bqone}{{B_{1q}}}

\newcommand{\mDoz}{m_{D_1^{0}}}
\newcommand{\mDod}{m_{D_1^{+}}}
\newcommand{\mDos}{m_{D_{1s} }}
\newcommand{\mBoz}{m_{B^{0}_1}}
\newcommand{\mBou}{m_{B^{+}_1}}
\newcommand{\mBos}{m_{B_{1s}}}

\newcommand{\Mbar}{{\bar{M}}}

\newcommand{\TAB}{Tab.~}
\newcommand{\TABs}{Tabs.~}
\newcommand{\FIG}{Fig.~}
\newcommand{\SEC}{Sec.~}

\newcommand{\APP}{App.~}
\newcommand{\APPs}{Apps.~}
\newcommand{\EQ}{Eq.~}
\newcommand{\EQs}{Eqs.~}

\newcommand{\ORD}{{\cal O}}
\newcommand{\MSbar}{\overline{\text{MS}}}

\newcommand{\fBst}[1]{f#1_{B^*}}
\newcommand{\fBone}[1]{f#1_{B_1}}

\newcommand{\err}[2]{\phantom{}^{+#1}_{-#2}}
\newcommand{\Verr}[3]{#1^{+#2}_{-#3}}

\newcommand{\qbarq}{\langle \bar{q}q\rangle}
\newcommand{\sbars}{\langle \bar{s}s\rangle}

\newcommand{\lscale}[1]{{ |#1\!\GeV}}

\newcommand{\Corr}{\Pi}

\newcommand{\muUV}{\mu_{\text{UV}}}
\newcommand{\mukin}{\mu_{\textrm{kin}}}
\newcommand{\muM}{\mu_{m}}
\newcommand{\muals}{\mu_{\alpha_s}}
\newcommand{\muCond}{\mu_{\text{cond}}}

\newcommand{\Bx}{B_{(i)}}

\newcommand{\fb}{f_{B}}
\newcommand{\fbs}{f_{B^*}}
\newcommand{\fbo}{f_{B_1}}
\newcommand{\fbsT}{f_{B^*}^T}
\newcommand{\fboT}{f_{B_1}^T}

\newcommand{\fbS}{f_{B_s}}
\newcommand{\fbsS}{f_{B_s^*}}
\newcommand{\fboS}{f_{B_{1_s}}}
\newcommand{\fbsTS}{f_{B_s^*}^T}
\newcommand{\fboTS}{f_{B_{1_s}}^T}

\newcommand{\fd}{f_{D}}
\newcommand{\fds}{f_{D^*}}

\newcommand{\fdsT}{f_{D^*}^T}

\newcommand{\fdS}{f_{D_s}}
\newcommand{\fdsS}{f_{D_s^*}}

\newcommand{\fdsTS}{f_{D_s^*}^T}

\newcommand{\sSum}{S(\underline{\alpha})\!+\!\tilde{S}(\underline{\alpha})}
\newcommand{\tSum}{\sum\limits_{{i=1}}^{4}T_i^{(1)}(\underline{\alpha})}

\newcommand{\logMu}[1]{\ln\left(\frac{\mu_{\text{#1}}^2}{m_b^2}\right)}

\newcommand\mc[1]{\multicolumn{1}{c|}{#1}} % handy shortcut macro 
\newcommand\mct[1]{\multicolumn{1}{c||}{#1}} % handy shortcut macro   
\newcommand\mcc[1]{\multicolumn{1}{c}{#1}} % handy shortcut macro   

\newcommand*{\mathcolor}{}
\def\mathcolor#1#{\mathcoloraux{#1}}
\newcommand*{\mathcoloraux}[3]{%
  \protect\leavevmode
  \begingroup
    \color#1{#2}#3%
  \endgroup
}

\begin{document}

\preprint{CP3-Origins-2020-13 DNRF90}
%\begin{flushright}  TTP20-019 \\ P3H-20-017 \end{flushright}%}

\title{\boldmath Radiative  Decays of Heavy-light Mesons  \\ and the $f_{H,H^*,H_1}^{(T)}$  Decay Constants 
}

\author[1]{Ben Pullin,}
\author[1]{Roman Zwicky}

\affiliation[1]{Higgs Centre for Theoretical Physics, School of Physics and Astronomy, University of Edinburgh, Edinburgh EH9 3JZ, Scotland}

\emailAdd{b.pullin@ed.ac.uk, roman.zwicky@ed.ac.uk}

\abstract{
The on-shell matrix elements,  or couplings $g_{H H^*(H_1)\gamma}$,  
describing the $B(D)_q^* \to B(D)_q \gamma$ and $B_{1 q} \to B_q \gamma$ ($q = u,d,s$)
radiative decays,  are determined
from light-cone sum rules at next-to-leading order for the first time. 
Two different interpolating operators are used for the vector meson, providing additional 
robustness to our results.
For the $D^*$-meson, where some rates are  experimentally known,  agreement is found. 
The couplings are of additional  interest as they govern the lowest pole residue in the 
 $B(D) \to \gamma$ form factors which in turn are connected to QED-corrections in leptonic decays $B(D) \to \ell \bar \nu$.  
 Since the couplings and residues are related by the decay constants
$f_{H^*(H_1)}$  and $f^T_{H^*(H_1)}$, we determine  them at next-leading order as a by-product. The quantities  $\{ f_{H^*}^T, f_{H_1}^T\}$  
have not previously been subjected to a QCD sum rule determination. 
All results are compared with the existing experimental and theoretical literature.}
\maketitle

\flushbottom

\setcounter{tocdepth}{3}
\setcounter{page}{1}
\pagestyle{plain}

\section{Introduction}

In this paper, we consider the on-shell couplings $g_{HH^*\gamma}$ 
and $g_{HH_1\gamma}$, 
for  $B(D)_q^* \to B(D)_q \gamma$ and $B_{1 q} \to B_q \gamma$ where $q = u,d,s$,   from
 light-cone sum rules (LCSR) \cite{Balitsky:1997wi,Colangelo:2000dp}.\footnote{\label{foot:D1}  For the $1^+$ state $H_1$ we only consider the $B_1$-state since the $D_1$-state 
is already overshadowed by the $D  \pi \pi$ 3-particle state ($m_{D_1} - m_{D} - 2 m_{\pi} \approx 270\MeV$). 
This effect is less pronounced, as a result of $m_c/m_b$ suppression,  for the $B_1$ since 
$m_{B_1} - m_{B}- 2 m_{\pi} \approx 160\MeV$.} 
Our own interest in these couplings is two-fold. Firstly they describe 
the decay $H^*(H_1) \to H \ga$; secondly they appear as residues of 
the $m^2_{H^*(H_1)}$-pole for the $H \to \ga$ form factor  e.g. \cite{Janowski:2021yvz}
and are likely dominant at the kinematic endpoint.  
The form factors in this kinematic region are of importance for 
QED-corrections to $H \to \ell \bar \nu$ and $H \to \ell \ell$. 
The neutral form factor is an ingredient for 
 the Standard Model prediction of $B_s \to \mu\mu \ga$  \cite{GRZ17,Kozachuk:2017mdk} and 
invisible particle searches in $B_s \to \ell\ell  X$ 
(where $X$ could be a  flavoured axion or a dark photon  
at the LHCb, CMS or ATLAS experiment  \cite{Albrecht:2019zul}).

The results derive from the same correlation  functions as the form factors but involve 
 a double, rather than a single, dispersion relation.  The additional 
 dispersion variable is the momentum transfer of the form factor $q^2$ where the $H^*,H_1$-meson is the lowest lying state. 
  This is a technically involved matter at next-to-leading order (NLO), and our computation provides the first complete NLO computation 
 at twist-$1$ and -$2$ level, utilising the master integrals from \cite{DiVita:2017xlr,Janowski:2021yvz}.  
 A notable aspect is that  the kinetic mass scheme  \cite{Bigi:1994em}, gives more stable results than the $\MSbar$- 
 and the pole-scheme. 
 
The residues and the couplings differ, apart from ratios of known hadron masses,   by decay constants (cf.  \SEC\ref{sec:defs}).   
We determine five distinct decay constants from local  QCD sum rules (SRs) \cite{SVZ79I,SVZ79II}
to ensure  consistency of our results; the well-known pseudoscalar $f_H$ and both the vector $f_{H^*}(f_{H_1}  ) $ and tensor  
$f_{H^*}^T(f_{H_1}^T)$ of the $1^-$($1^+$) state.
To the best of our knowledge 
$\{ f_{H^*}^T, f_{H_1}^T\}$  have not previously been determined from QCD SRs.
A relevant feature is that some $D^*$ couplings are known from experiment.
This is not the case for  the $B^*$ as the unknown total width means that the coupling values cannot be inferred.

The $g_{H^*H\ga}$ couplings have been considered in LCSR to LO
in \cite{Aliev:1995zlh} and at NLO at twist-$2$ level \cite{Li:2020rcg}. 
Lattice determinations of $g_{D^*D \ga}$ (with large uncertainty) \cite{Becirevic:2009xp} 
and $g_{D_s^*D_s \ga}$ (with small uncertainty) \cite{Donald:2013sra} are available. 
Heavy-light meson decay constants have been evaluated  to NLO (and partially beyond) 
in  \cite{Jamin:2001fw,Gelhausen:2013wia,Wang:2015mxa} in SR. Lattice results are numerous and 
 include \cite{Becirevic:2012ti,Lubicz:2017asp}.

The paper is organised as follows. In \SEC\ref{sec:defs} we define the couplings and give 
their relations to the residues of the form factors. 
\SEC\ref{sec:comp} is concerned with the main SR aspects of the couplings e.g.
the computation,  the double dispersion 
relation and the Borel transform (with more detail in  \APPs\ref{app:double} and 
\ref{app:double-borel}). The main results for the residues and the couplings 
 are given in \TABs \ref{tab:rValues} and \ref{tab:gValues}  respectively. 
 The decay constants, as bona fide predictions, 
are presented in  \SEC\ref{sec:decaySR}, with  analytic results 
in  \APP\ref{app:decayConsts}.
Numerical values of decay constants and ratios thereof are collected in  \TABs\ref{tab:fBparams} and \ref{tab:fBRatio} respectively.
We conclude in \SEC\ref{sec:conc}.
Conventions, definitions  and inputs are grouped into \APP\ref{app:conv}.

\section{The Couplings \texorpdfstring{$g_{HH^*(H_1)\ga}$}{} and their Relation to \texorpdfstring{$H \to \ga$}{} Form Factors }
\label{sec:defs}

The purpose of this section is to discuss relevant method-independent aspects
of the computation. For concreteness we shall write $H = B$, throughout this section, 
which stands for either of the beauty $B_{u,d,s}$- or  charmed  $D_{u,d,s}$-mesons.
The couplings of interest are defined from the on-shell 
 amplitudes\footnote{More concretely the couplings parametrise the on-shell matrix elements $\langle  \bar B(p_B) \ga(k) | \bar B^*(q) \rangle   = [ - i (2\pi)^4 \de^{(4)} (\sum p_i)]  \, {\cal A}_{B^* \to B \ga}$ and $\langle  \bar B(p_B) \ga(k) | \bar B_1(q) \rangle   = [ - i (2\pi)^4 \de^{(4)} (\sum p_i)]  \, {\cal A}_{B_1 \to B \ga}$. }
\begin{alignat}{1}
  {\cal A}_{B^* \to B \ga}   \;=\;&  \frac{i}{2} s_e   e \,\varepsilon^{\al\be\ga\de}(p_B)_{\al}\,\eta_{\be}\,F_{\ga\de}\,   \, \gonemi   \;,\nonumber \\[0.1cm]
  {\cal A}_{B_1 \to B \ga}   \;=\;&  - s_e  e F_{\al \be} (p_B)^\al \eta^\be  \,  \gonepl  \;,
\end{alignat}
where  $D_\mu = \partial_\mu + i e s_e A_\mu$ (with $s_e = \pm 1$ depending on convention),
$\eta$ is the vector meson's polarisation, $F_{\al \be}  = i k_{[\al} \eps^*_{\be]}$ stands for 
the photon's outgoing plane wave and the  coupling's mass dimension  is 
$[\gonemi]=[\gonepl]=-1$.
 We refer the reader to \APP\ref{app:conv} for more details on conventions.
For the decay rates, 
with $\al=e^2/4\pi$ as  the fine structure constant, we obtain
\begin{align}
\label{eq:widths}
\Gamma(B^*\to B\gamma)&=\frac{\al}{24}   \left( 1-\frac{m_B^2}{m_{B^*}^2} \right)^3 m_{B^*}^3 \gonemi^2\;,\nonumber \\[0.1cm]
\Gamma(B_1\to B\gamma)&= \frac{\al}{24} 
\left( 1-\frac{m_B^2}{m_{B_1}^2} \right)^3 m_{B_1}^3 
\gonepl^2\;,
\end{align}
where the first expression  agrees with \cite{Amundson:1992yp} for example.
These rates follow from an 
 effective Lagrangian of the form\footnote{One might wonder whether the proximity of the $B$ and $B^*$ mass leads to any enhanced terms in the soft photon region in diagrams where the photon couples to an external $B$-meson and a lepton for instance 
 (e.g. diagram top left of figure 3 in \cite{Isidori:2020acz} where the weak Hamiltonian corresponds to
 $B^* \to K \ell \ell$). 
  The behaviour of the denominator in the soft region (i.e. $k_\mu \to 0$),  $\frac{1}{ 2 k \cdot p_B + \Delta M^2_B}
  \frac{1}{k \cdot \ell_1}\frac{1}{k^2}$ with $\Delta M^2_B  = m_{B^*}^2- m_B^2 = \ORD(m_b \Lambda_{\textrm{QCD}})$,  is softened by the derivative term $F_{\al\be}$ and avoids  
 unsuppressed large logarithms of the form $\ln( \Delta M^2_B/m_B^2)$. This is another manifestation, with a different twist,  of the finding in  \cite{Isidori:2020acz} (cf. section 3.4 therein)   that structure dependent terms do not generate large logarithms.}
\begin{equation}
\label{eq:Leff}
{\cal L}_{\textrm eff} =  s_e  \gonemi \, \frac{1}{2}\epsilon(B^*,\partial B^\dagger,F) - i s_e  \gonepl \, B^*_\al \partial_\be B^\dagger F^{\al \be}  + \textrm{h.c.} \;.
\end{equation}
 This  Lagrangian can be used at small recoil 
and has to be supplemented by higher order couplings away from it.

As mentioned earlier, another point of interest in the couplings arises from their relation to 
pole-residues of the $\bar{B} \to \ga$ form factors \cite{Janowski:2021yvz} (and cf. \APP\ref{app:conv}).\footnote{We note that when translating between the $\bar{B}\to\gamma$ and $B\to\gamma$ form factors only the axial, and not the vector parts change sign, as can be inferred by applying a charge $C$-transformation with $C|\bar B\rangle=|B\rangle$. We stress that our results are formally quoted for the $\bar{B}$-meson.} 
For clarity let us consider the dispersion representation of the vector form factor 
\begin{equation}
\label{eq:VFF}
V^{\bar{B} \to \ga}_{\perp[ \parallel]} (q^2) =   
\frac{1}{\pi} \int^\infty_{\textrm{cut}}  dt \frac{\Ima[ V^{\bar{B} \to \ga}_{\perp[ \parallel]} (t)]}{t - q^2 - i0} =  
\frac{r^{V}_{\perp[ \parallel]}}{1 - q^2/m_{B^*[B_1]}^2}  + \dots \;,
\end{equation}
where the dots represent  higher terms in the spectrum.
For the tensor form factor, $T^{\bar{B} \to \ga}_{\perp[ \parallel]} (q^2)$,  
the analogous form holds.
The relation of the  residues to the couplings are 
\begin{alignat}{4}
\label{eq:r}
& r^{V}_{\perp}     & \;=\; & \frac{m_B \fBst{} }{m_{B^*}} \gonemi   \;, \qquad & & r^{T}_{\perp} (\muUV)    & \;=\; & \fBst{^T} (\muUV) \gonemi \;,\nonumber \\[0.1cm]
& r^{V}_{\parallel} & \;=\; &   \frac{m_B \fBone{} }{m_{B_1}} \gonepl \;,\qquad    & & r^{T}_{\parallel}(\muUV) & \;=\; &   \fBone{^T}(\muUV) \gonepl \;,
\end{alignat}
with decay constants $f^{(T)}_{B^*(B_1)}$ defined  in \eqref{eq:decay_const}.
The following exact relations, with $\muUV$-dependence suppressed,  
\begin{equation}
\label{eq:test}
 \frac{r^{V}_{\perp}}{ r^{T}_{\perp} } = \frac{m_B \fBst{} }{m_{B^*}\fBst{^T} } \;, 
 \quad \frac{r^{V}_{\para}}{ r^{T}_{\para} } =  \frac{m_B \fBone{} }{m_{B_1}\fBone{^T}} \;,
\end{equation}
are a consequence of the freedom to choose a particle's interpolating operator in field theory. 
This provides us with a non-trivial consistency check of our SR evaluation.
 Finally a note on the ultraviolet (UV)  scale dependence $\muUV$. 
 The couplings are of course scale-independent since they correspond 
 to on-shell matrix elements. Thus the vector residues are scale-independent whereas 
the tensor ones scale like the tensor decay constant 
\begin{equation}
\label{eq:gaT}
\ga_T = - \frac{d}{d\ln \muUV} \ln f^T_{B^*(B_1)}(\muUV) =  \com{  - \frac{d}{d\ln \muUV} \ln r^T_{\perp(\para)}(\muUV) }    \;, \quad 
 \end{equation} 
 with 
\begin{equation}
 \ga_T =  \frac{\al_s}{4 \pi} 2 C_F  + O(\al_s^2) \;,
\end{equation} 
 and
 $C_F = (N_c^2-1) /(2 N_c) = 4/3$.

\section{The  \texorpdfstring{$g_{HH^*(H_1)\gamma }$}{}  Couplings from Light-cone Sum Rules}
\label{sec:comp}

\subsection{The Computation}

The couplings can be computed within the framework of QCD SRs on the light-cone. Proceeding via standard techniques we define two correlation functions \cite{Janowski:2021yvz}
\begin{align}
\label{eq:corr}
\Corr^\type_{\perp\mu}(p_B,q) & \; \equiv \;  i \int_x  e^{-ip_B \cdot x} \matel{\ga(k,\eps)}{ T J_{B_q}(x) O_{\perp\mu}^\type(0)}{0}\;=\;s_e  \Pperp{\mu} \, \Corr^\type_\perp(p_B^2,q^2)  \;,\\[0.2cm]
\Corr^\type_{\parallel\mu}(p_B,q) & \; \equiv \;  i \int_x  e^{-ip_B \cdot x} \matel{\ga(k,\eps)}{ T J_{B_q}(x) O_{\parallel\mu}^\type(0)}{0}\;=\;s_e \left(  - \Ppara{\mu} \, \Corr^\type_\parallel(p_B^2,q^2) + \dots\right) ,\nonumber
\end{align}
with quantum numbers chosen such that $\Corr^\type_\perp$ and $\Corr^\type_\parallel$ contain information on $\gonemi$ and $\gonepl$, respectively and $\Gamma\in\{V,T\}$. Above the shorthand  $\int_x = \int d^4 x$ has been adopted and the dots represent structures  \cite{Janowski:2021yvz} which are not important for this discussion. The $B$-meson is interpolated by the operator $J_{B_q}$ 
\begin{equation}
\label{eq:fBjB}
J_{B_q} \equiv (m_b + m_q) \bar{b} i \ga_5 q \;, \quad 
 \matel{\bar{B}_q}{J_{B_q}}{0} = m_{B_q}^2 f_{B_q} \;,
\end{equation}
and the Lorentz structures $\Pperppara{\mu}$ are given by
\begin{equation}
\label{eq:P} 
 \Pperp{\mu} \equiv   \varepsilon_{\mu \rho \be \ga } \eps^{*\rho}(p_B)^\be k^\ga  \;,  \quad    \Ppara{\mu}   \equiv  i  \,  (   p_B \!\cdot \!k \, \eps^*_{\mu} -   \,p_B \!\cdot \!\eps^*\,k_{\mu} ) \;.
\end{equation}
with $\eps$  the  photon's polarisation vector, $p_B=q+k$ and on-shell momentum  $k^2 = 0$. The vector and tensor operators of the $b\to q$ effective Hamiltonian are given by
\begin{align}
  O^V_{\perp\mu} &\equiv -  \frac{1}{e} m_{B_q} \bar{q} \gamma_{\mu} b \;, \qquad O^T_{\perp\mu}  \equiv \frac{1}{e}  \bar{q} i q^\nu \sigma_{\mu \nu} b \;,\nonumber\\[0.1cm]
  O^V_{\parallel\mu} &\equiv   \frac{1}{e} m_{B_q} \bar{q} \gamma_{\mu}  \ga_5 b \;, \qquad O^T_{\parallel\mu}  \equiv \frac{1}{e}  \bar{q} i q^\nu \sigma_{\mu \nu} \ga_5 b \;.
\end{align}
For brevity, from this point onwards we drop the subscript denoting the quark flavour such that $m_{B_q}\!=\!m_B$, $f_{B_q}\!=\!f_B$, et cetera.
As previously 
mentioned, the computation  of the correlation function is the same
as for the $\bar{B} \to \ga$ form factor; we refer the reader to \cite{Janowski:2021yvz} for details of the calculation and now turn  to the double dispersion relation.

\subsection{The Dispersion Relation}

The hadronic representation of the correlation functions is obtained from the double 
discontinuity of the correlation function\footnote{\label{foot:schwartz}
As Schwartz's reflection principle applies, one may use $\text{Disc}\rightarrow 2i\,\text{Im}$ cf. \cite{Zwicky:2016lka} for instance.} 
\begin{equation}
\rho_{\text{had}^*}^{\Gamma}(p_B^2,q^2) = \frac{1}{(2 \pi i)^2 } \disc_{q^2} \disc_{p_B^2}   \Pi^\Ga_\perp(p_B^2,q^2) \;,
\end{equation}
and reads
\begin{align}\label{eq:corr_hadron}
s_e \Corr_{\perp\mu} ^{\Gamma} &= \sum_{\textrm{pol}} \frac{ \matel{0}{J_B}{\bar B} {\cal A}_{B^* \to B \ga} \matel{\bar B^*}{O_{\perp\mu}^\Gamma}{ 0}}{(m_{B^*}^2-q^2)(m_B^2-p_B^2)} + P^{\perp}_{\mu}\int\!\!\!\!\int_{\Sigma_\perp}      ds\, dt\frac{\rho_{\text{had}^*}^{\Gamma}(s,t)}{(t-q^2)(s-p_B^2)}+\dots,\nonumber \\[0.1cm] 
s_e \Corr_{\parallel\mu} ^{\Gamma} &= \sum_{\textrm{pol}} \frac{ \matel{0}{J_B}{\bar B} {\cal A}_{B_1 \to B \ga} \matel{\bar B_1}{O_{\para\mu}^\Gamma}{ 0}}{(m_{B_1}^2-q^2)(m_B^2-p_B^2)} - P^{\parallel}_{\mu}\int\!\!\!\!\int_{\Sigma_\para}   ds\, dt\frac{\rho_{\text{had}_1}^{\Gamma}(s,t)}{(t-q^2)(s-p_B^2)}+\dots, 
 %\matel{0}{O_{\parallel\mu}^\Gamma}{ \bar B_1} {\cal A}_{B_1 \to B \ga}^*\matel{\bar B}{J_B}{0}
\end{align}
where the sum runs over the vector meson's polarisations. 
The integration domain $\Sigma_{\perp,\para}$ ranges from a lower cut shifted by two pion masses from the poles up to infinity.  
The dots indicate single dispersion integrals which do not contribute to the final result,
and can be seen as the analogues of the subtraction terms 
of single dispersion integrals. 

The matrix elements to the right are the decay constants
\begin{alignat}{6}\label{eq:decay_const}
&\matel{0}{(O_{\perp\mu}^V)^\dagger}{\bar{B}^*}&\;=\,&-\frac{1}{e}m_B m_{B^*}\fBst{}\eta_{\mu}\;, &\qquad& \matel{0}{(O_{\perp\mu}^T)^\dagger}{\bar{B}^*}&\;=\;&-\frac{1}{e}m_{B^*}^2f_{B^*}^T\eta_{\mu}\;,\nonumber \\[0.1cm]
&\matel{0}{(O_{\parallel\mu}^V)^\dagger}{\bar{B}_1}&\,=\;&\frac{1}{e}m_B m_{B_1}\fBst{}\eta_{\mu}\;, &\qquad &\matel{0}{(O_{\parallel\mu}^T)^\dagger}{\bar{B}_1}&\;=\;&\frac{1}{e}m_{B_1}^2f_{B_1}^T\eta_{\mu} \;,
\end{alignat}
 where $\eta$ is the vector mesons' polarisation vector e.g. \EQ\eqref{eq:fBs}.
 The SR procedure involves the Borel transformation in both variables, 
 $\Corr_{\|[\perp]}^{\Gamma}(M_1^2,M_2^2)\equiv\mathcal{B}_{M_2^2}^{q^2}\mathcal{B}_{M_1^2}^{p_B^2}\Corr_{\|[\perp]}^{\Gamma}(p_B^2,q^2)$, to enhance convergence. 
 In the case of a dispersion relation of the form \eqref{eq:corr_hadron} this is straightforward due 
 to the well-known formula
 \begin{equation}
 \label{eq:Borel-easy}
 \mathcal{B}_{M^2}^{q^2} \left(  \frac{1}{m^2-q^2} \right) = e^{-m^2/M^2} \;.
\end{equation} 
We refer the reader to \APP\ref{app:double-borel} for 
the  definition of the Borel transformation.

\subsection{The Light-cone Operator Product Expansion}
\label{eq:LCOPE}

 The correlation functions \eqref{eq:corr} are evaluated with perturbative QCD using 
 the light-cone operator product expansion (LC-OPE) ordered, in practice, by a converging expansion in twist. 
 The twist, known from deep inelastic scattering, is the dimension of the operator minus its spin. 
 We refer to \cite{Janowski:2021yvz} for specific details and to the technical \cite{Braun:2003rp} and applied \cite{Colangelo:2000dp} reviews on the subject. It seems worthwhile to state that, contrary  to intuition, the photon is more involved than an ordinary 
 vector meson as it has both  
 perturbative  (twist-$1$) \emph{and} non-perturbative nature (higher-twist).
The latter is encoded in the photon distribution amplitude (DA)
which can be understood as $\rho/\omega$-$\ga$ or $\phi$-$\ga$ conversions.
 At LO in $\al_s$ we perform the computation up to twist-$4$ including $3$-particle DAs,  whilst at next-to-LO (NLO)  twist-$1$ and twist-$2$ contributions have been computed.  
 See however \SEC\ref{sec:numg} for remarks on the completeness of twist-$4$.

 \subsubsection{The ``Partonic" Dispersion Relation} 
 
 One may also write a dispersion relation in perturbative QCD, 
 \begin{equation}
\label{eq:double_disp}
\Corr_{\parallel[\perp]}^{\Gamma}(p_B^2,q^2)=\int_{m_b^2}^\infty ds\int_{m_b^2}^\infty
 dt\frac{\rho^{\Gamma}_{\parallel[\perp]}(s,t)}{(t-q^2)(s-p_B^2)} + \dots
\end{equation}
 which is formally distinct by its slightly different analytic structure with the discontinuity  starting 
 at  $m_b^2$.\footnote{The $m_{u,d,s}$ masses are considered 
 in the linear approximation for which we have derived new results 
 such as the $m_q$-correction to the twist-$2$ photon DA \cite{Janowski:2021yvz}.}
 The dots have the same meaning as for the ``hadronic" dispersion relation.
  
 Performing the double dispersion relation at NLO is complicated by pole 
 singularities in $q^2 = p_B^2$. Taking a single discontinuity, say in $p_B^2$, one is faced 
 with 
 \begin{equation}
  \disc_{p_B^2}   \Pi^\Ga_\perp(p_B^2,q^2)  = \sum_{n=0}^{3 }\frac{\rho_i (q^2,p_B^2)}{(q^2 -p_B^2)^n} \;,
 \end{equation}
 where the $\rho_i$ themselves contain non-trivial cuts.\footnote{At LO this is not the case and this is what makes 
 them considerably  easier to handle in practice.}
  These singularities, dubbed second type singularities \cite{Itzykson:1980rh, Zwicky:2016lka}, are solutions of the Landau equation for 
  $\Pi^\Ga_\perp(p_B^2,q^2)$ but are not on the physical sheet. However this changes once the discontinuity 
  is taken in $p_B^2$ and they need to be taken into account. 
  We refer the reader to \APP\ref{app:double} for technical details.\footnote{An alternative 
  is to use Schwartz's reflection principle, $\text{Disc}\rightarrow 2i\,\text{Im}$, to obtain the discontinuity cf.
  footnote \ref{foot:schwartz}. 
  One can then deform the $dt$-integration path into the complex plane, away from the poles, in order to obtain 
  a working dispersion  representation. This approach, whilst being computationally inefficient, provides numerically stable results as long as the upper integration boundaries in the $dt$ and $ds$ integrals are sufficiently far apart. However, given the almost degenerate values of the masses $m_B$ and $m_{B^*}$ a sufficient separation of the upper boundaries can not be justified, rendering this approach sub-optimal.
} 
  
  \subsubsection{Borel Transformation of LO Terms for generic Distribution Amplitudes} 
  
  As previously stated, for a given dispersion representation \eqref{eq:double_disp} the Borel transformation 
  is straightforward due to \eqref{eq:Borel-easy}. However, this demands committing to a specific 
  DA. 
 As these can improve over time, due  to better determination of hadronic parameters,
  there is some advantage in keeping them generic. Let us consider
  \begin{equation}
 \label{eq:F0}
  \Pi(p_B^2,q^2)  \supset  \int_0^1 du \frac{(q^2)^{\ell}f_{n}(u)}{(m_b^2 - u p_B^2 - \bar u q^2)^n} \;, 
  \quad \ell = 0,1 \;, \quad n = 1,2,3 \;,
 \end{equation}
 where  $f_n(u)$ is some function proportional to the DA with suitable features 
 in order to be compatible with first principle analytic properties.
 How to perform the Borel transformation \textit{and} the continuum subtraction 
 is described in \APP\ref{app:double-borel}. These results extend those currently seen in the literature and are presented in greater detail.
 In theory a double Borel transform provides two Borel parameters. In practice however, we content ourselves to setting them equal 
 \begin{equation}
 \label{eq:BorelM2}
 M_1^2=M_2^2=2 \Mbar^2 \to 2 \hat{M}^2  \;,
 \end{equation}
  (and $u_0=1/2$ cf.  \eqref{eq:u0Mhat}), which is 
 justified since $m_B \approx m_{B^*} (\approx m_{B_1})$.  The $3$-particle DAs can be handled with the 
 same technique  as they reduce to an effective
 $2$-particle DA (cf. appendix D in reference \cite{Janowski:2021yvz}).

\subsection{The Sum Rule}

The final step in completing the SR is to invoke 
semi-global quark-hadron duality.
For a double dispersion relation this is not straightforward.  Before addressing this 
issue let us assume an integration region (parametrised by a single parameter $a$ 
and $\bar\de^{(a)}_{s,t}$ specified in the next subsection), 
implemented with step function on the 
spectral density
\begin{equation}
\label{eq:step}
\rho_{\text{had}_1[\text{had}^*]}^{\Gamma}(s,t)=\rho^{\Gamma}_{\parallel[\perp]}(s,t)\Theta(s-\bsup{m_b^2} )\Theta(t-\btup{s} )\;.
\end{equation}
Equating the``partonic" and ``hadronic" parts one obtains the sum rule
\begin{align}\label{eq:rSR}
f_B r^{\Gamma}_{\parallel[\perp]}=\frac{1}{m_B^2 m_{B_1[B^*]}^2}\int_{m_b^2}^{\bsup{m_b^2}} ds\int_{m_b^2}^{\btup{s}} dt \,e^{\frac{m_B^2 -s}{2\hat{M}^2} }
e^{\frac{m_{B_1[B^*]}^2 -t}{2\hat{M}^2} }\rho^{\Gamma}_{\parallel[\perp]}(s,t)\;,
\end{align}
with the relation between the couplings and the residues $r^{\Gamma}_{\parallel[\perp]}$ given in \eqref{eq:r}
and $\Gamma = V,T$.  
The somewhat unconventional factor of two in the exponent is a consequence of our definition of the Borel mass 
\eqref{eq:BorelM2}.
The LCSR determines the product $f_B r^{\Gamma}_{\parallel[\perp]}$ and to obtain 
the residues and the couplings one replaces the decay constants by a QCD SR to the same 
accuracy in $\al_s$, e.g. 
\begin{equation}
\label{eq:recipe}
r^{\Gamma}_{\parallel[\perp]} = \frac{ [f_B r^{\Ga}_{\para[\perp]} ]_{\textrm{LCSR}} } {[f_B]_{\textrm{SR}}} \;, 
\end{equation}
and
\begin{alignat}{3}
\label{eq:recipe2}
&g_{BB^*\ga} &\;=\;& \frac{m_B^*}{m_B} \frac{ [f_B r^{V}_{\para[\perp]} ]_{\textrm{LCSR}} } {[f_B]_{\textrm{SR}}[f_{B^*}]_{\textrm{SR}}} &\;=\;& 
 \frac{ [f_B r^{T}_{\para[\perp]} ]_{\textrm{LCSR}} } {[f_B]_{\textrm{SR}}[f^T_{B^*}]_{\textrm{SR}}} 
  \;, \nonumber \\[0.1cm] 
& g_{BB_1\ga} &\;=\;& \frac{m_{B_1}}{m_B}  \frac{ [f_B r^{V}_{\para[\perp]} ]_{\textrm{LCSR}} } {[f_B]_{\textrm{SR}}[f_{B_1}]_{\textrm{SR}}} &\;=\;&   \frac{ [f_B r^{T}_{\para[\perp]} ]_{\textrm{LCSR}} } {[f_B]_{\textrm{SR}}[f^T_{B_1}]_{\textrm{SR}}}
\;.
\end{alignat}
As previously mentioned, the two determinations for each couplings serve as an additional quality test of our SR.

\begin{figure}[]
\centering
  \begin{minipage}{0.5\textwidth}
    \centering
    \includegraphics[width=1.\textwidth]{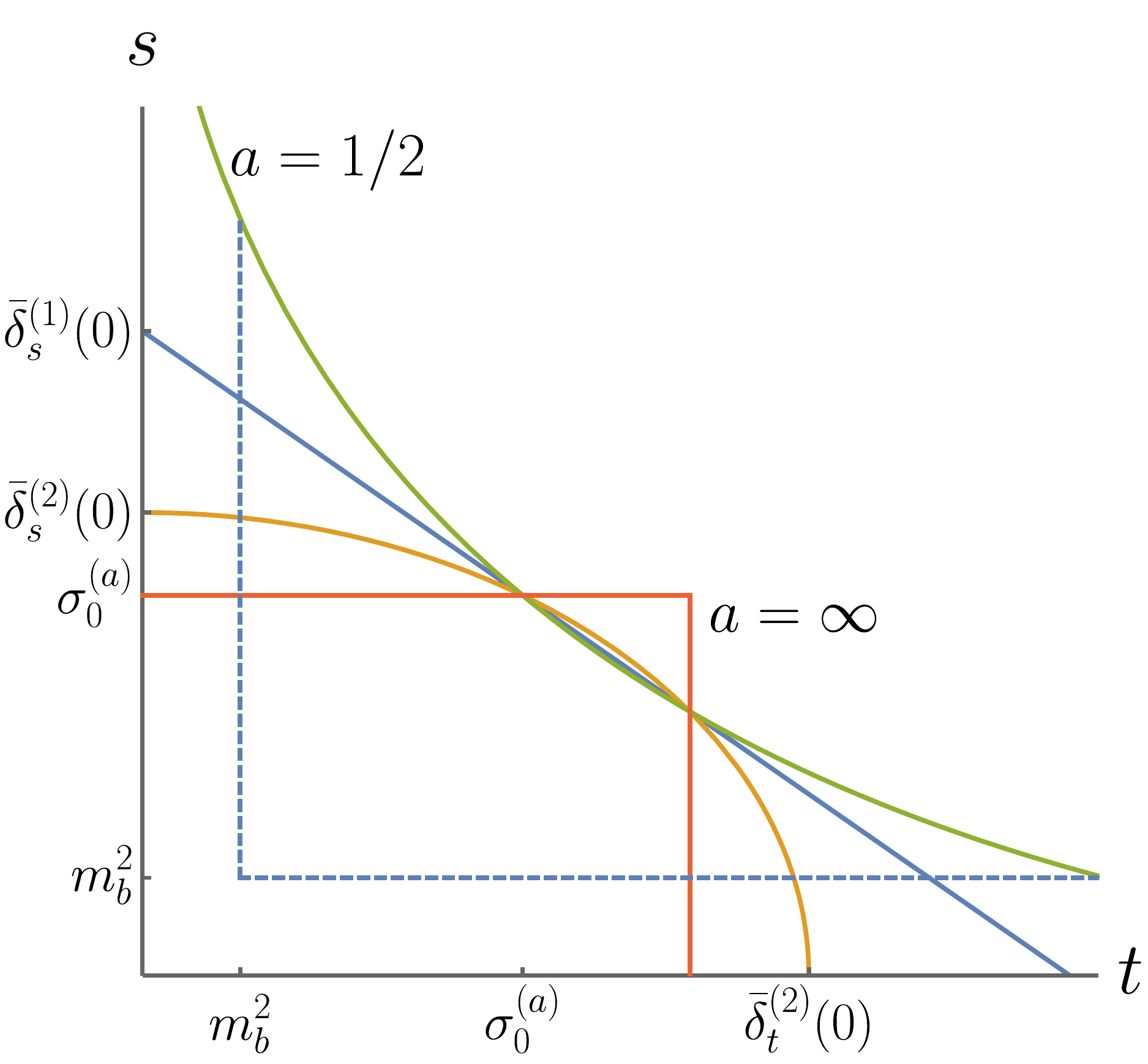}
  \end{minipage}\hfill
  \begin{minipage}{0.5\textwidth}
    \centering
    \includegraphics[width=1.\textwidth]{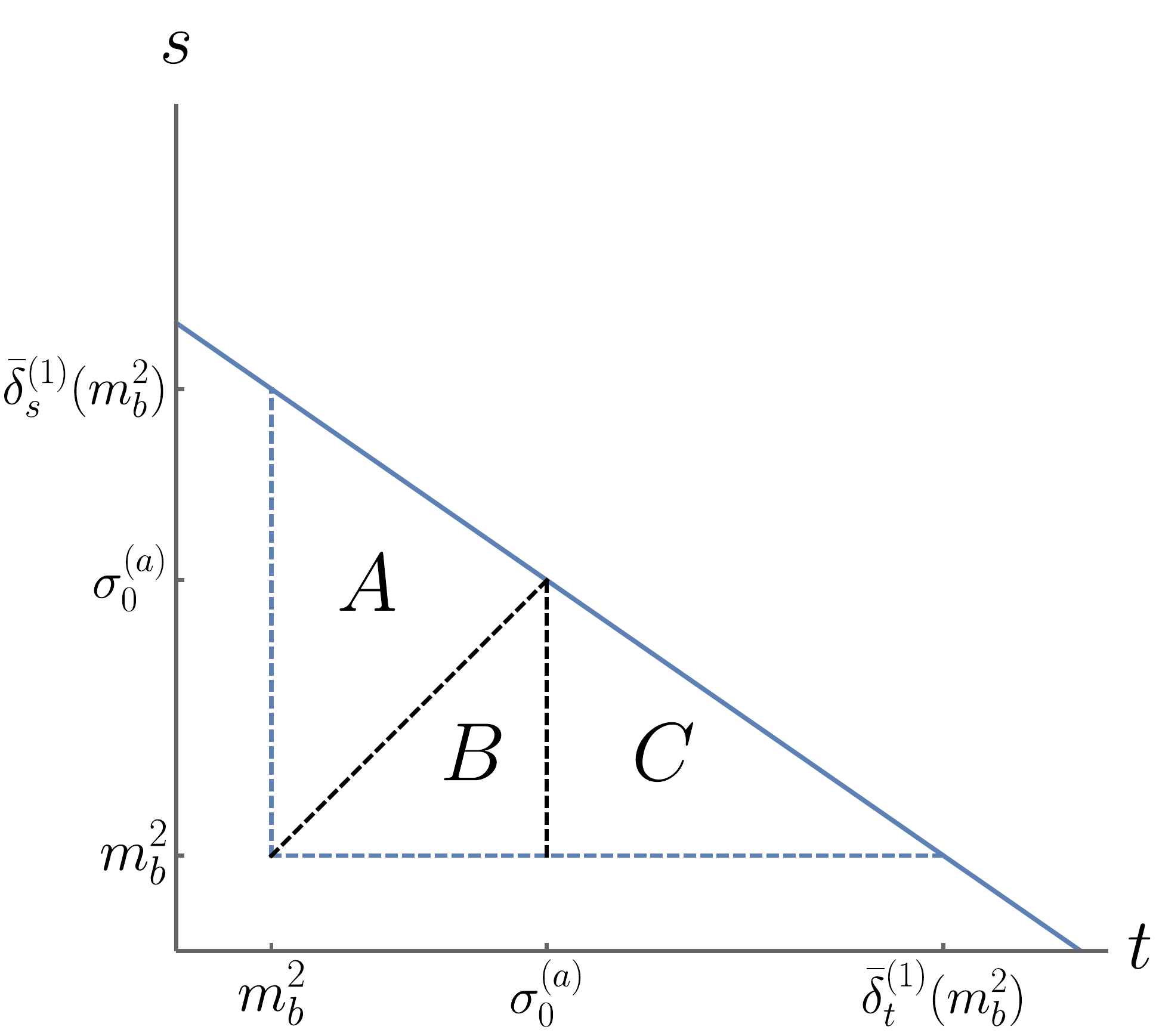}
  \end{minipage}
\caption{An overview of the duality interval. The left hand figure demonstrates how the parameterisation \eqref{eq:t0}, keeping the quantity $\Sred$ fixed, leads to a range of possible duality windows depending on the value of the parameter $a$. The solid green, blue, yellow, and orange curves correspond to the $a=1/2,1,2,\infty$, cases respectively. In the limit $\szerot\to \tzerot$ the curves intersect at a single point, $\Sred$. The right hand plot provides a more detailed view of the case $a=1$, which we adopt for our evaluation of the couplings. We note that the choice of duality window has little impact on the final result, cf \TAB\ref{tab:gValuesA}. The dashed blue line indicates the lower boundary on the duality window, enforced by the restriction that both $t$ and $s$ can only take values above the cut starting at $m_b^2$. The dashed black line indicates a technical division of the duality region necessary for application of the principal part prescription, cf. \eqref{eq:PrincPart}.}
\label{fig:regions}
\end{figure}
\subsubsection{Duality Region as a Function of the Duality Parameter \texorpdfstring{$a$}{}}
\label{sec:duality}

Finally we turn to the question of the duality region encoded in \eqref{eq:step}
and derive explicit relations as a function of the parameter $a$.
In defining the duality region,
\begin{equation}
\label{eq:t0}
\left( \frac{s}{\szerot } \right)^a + \left( \frac{t}{\tzerot} \right)^a   \leq  1  \;, 
\end{equation}
we follow earlier work \cite{Balitsky:1997wi,Khodjamirian:2020mlb} but  extend it in that we consider $\szerot, \tzerot$ as a function of the parameter $a$.
The solutions to the boundary defined by \eqref{eq:t0}, and which therefore enter \eqref{eq:step}, are
\begin{equation}\label{eq:dualParams}
 \bsup{t} =   \szerot \left(1 - \left( \frac{t}{\tzerot}\right)^a \right)^{1/a} \,,  \quad 
 \btup{s}  = \tzerot \left(1 - \left( \frac{s}{\szerot}\right)^a\right)^{1/a}  \,.
\end{equation}
A further quantity that arising from the parameterisation, and thus appearing in results given in the appendix, is 
\begin{equation}
\label{eq:Sred}
\Sred =  \frac{\szerot \tzerot }{ (\szerot^a +\tzerot^a)^{1/a} }  \;.
\end{equation}
Its geometric meaning can be inferred from \FIG\ref{fig:regions}. It 
takes on the r\^ole of the  single dispersion effective threshold if $\rho^\Gamma \propto \de(s-t)$ which
is the case for a large part of the contributions. Fortunately, variation of the duality parameter $a$ does not lead to large effects
when the daughter sum rule is invoked to constrain the SR parameters, as will be discussed in the next section. 

We turn to the question of which choice of the parameter $a$ is suitable.  We find that in the majority of cases the dependence of the couplings on the duality window is rather limited, as evidenced by \TAB\ref{tab:gValuesA}. The exceptions are the   $B_{1s}$- and the $D^*_{s}$-meson cases, showing more significant variation. 
\com{It has been argued that for the Isgur-Wise function \cite{Neubert:1991sp}
and the small velocity limit \cite{Blok:1992fc} that $a=1$ is a necessary choice. 
Whether or not this translates to other cases and in particular to the case at hand is an open question.}  We  adopt $a=1$ as our default choice, and include variations under the duality window in our estimate of the total uncertainty (cf. \SEC\ref{sec:gcomp}).

\subsection{Numerical Analysis}
\label{sec:numg}

Physical input parameters used for the numerical evaluation of the SRs can be found in \TAB\ref{tab:inputParams} in the appendix.

As there are a number of different renormalisation scales involved we 
discuss them in some detail.
The UV scale, $\muUV$, has already been mentioned  
below \EQ\eqref{eq:test} and is set  to the pole mass $m_b(m_c)^\textrm{pole}$.
 For the LCSR there remains the scale of the coupling $\mu_{\al_s}$, 
the mass $\mu_m$ (or $\mukin$ cf. below) and the LC-OPE factorisation scale $\mu_F$. 
We set \com{$\mu_F^2 =m_B^2-(m_b^{kin}(1\GeV))^2 
(=m_D^2-m_c^2(m_c))$} which is a standard albeit not a necessary choice 
and equate $\mu_{\al_s} = \mu_F$. 
The choice of a mass scale is linked with a choice of mass scheme.  
For the $B\to\gamma$ form factors we have found \cite{Janowski:2021yvz} that 
the $\MSbar$- and the  pole-scheme give rise to large  effects in either 
  higher twist or at $\ORD(\al_s)$ rendering  both of them suboptimal.  
 \com{ For the $g_{BB^*(B_1)\ga}$ couplings the evidence for adopting the kinetic- over the $\MSbar$-scheme is less compelling (smaller improvement in twist-convergence). However, in an effort to remain consistent with our previous work \cite{Janowski:2021yvz}, we choose to adopt the kinetic-scheme for the evaluation of both the FF residues and the effective couplings.}
As the kinetic mass scheme, originally devised  for the inclusive decay operator product expansion (OPE) 
\cite{Bigi:1994em},  \com{can be considered as a compromise between the
$\MSbar$- and the 
pole-scheme. 
Moreover, it is indeed found that the kinetic scheme is stable under scale variation.}
The kinetic scale is set to $\mukin\!=\!1\GeV$, with further details in \cite{Janowski:2021yvz}.
 For the $D$-meson decays the situation is different and the $\MSbar$ scheme gives more stable results 
 than the kinetic scheme and we thus employ the $\MSbar$ scheme with the standard choice 
  $\mu_m\!=\!m_c(m_c)$. This might not come as a
 surprise since $m_c$ itself is closer to $\mu_{F}$ as compared to the $B$-case. 
 
As indicated in \EQs\eqref{eq:recipe} and \eqref{eq:recipe2}, to obtain the physical 
quantities one needs to divide by the decay constant(s) to the same order 
(cf. \SEC\ref{sec:decaySR} for their discussion). 
The new inputs are the condensates, given in \TAB\ref{tab:inputParams},  
 and the factorisation scale of the local OPE, denoted by $\muCond$, which is set to 
 $\muCond = \mu_F$ in order to facilitate cancellations in the ratio. 
 A summary of all renormalisation scales is given in \TAB\ref{tab:scales} (left).  
 Another aspect is that we drop twist-$4$ corrections, other than the pure quark condensates,  
as they are incomplete (requiring the inclusion of 
$4$-particle DAs \cite{Janowski:2021yvz}). The resulting uncertainty ought to be captured, at least in part,  by the variation of the Borel parameter. 
 
The SR parameters $\{\Sred,\hat{M}^2\}$ and $\{s^{f_B}_0, M^2_{f_B}\}$ 
are determined by a number of constraints. 
As usual the Borel mass is determined subject to two competing factors, contamination
 from higher states is effectively  suppressed by a small $\hat{M}^2$, whilst fast convergence of the LC-OPE favours a large $\hat{M}^2$ as higher terms in the expansion are accompanied by ever increasing inverse powers of the Borel mass.  The compromise of these two criteria, 
 resulting in an approximately flat 
 curve, is known as the Borel-window. To constrain the effective thresholds $\{ \szerot,\tzerot \}$
 the, formally exact, daughter SR for the sum of meson masses \eqref{eq:daughterSR} is employed
\begin{equation}\label{eq:daughterSR}
\mB^2+m_{B_1[B^*]}^2= {2\hat{M}^4}\frac{ d}{ d\hat{M}^2} \ln 
\int_{m_b^2}^{\bsup{m_b^2}} ds\int_{m_b^2}^{\btup{t}} dt\,
e^{-\frac{s+t}{2\hat{M}^2}} \, \rho^{\Gamma}_{\parallel[\perp]}(s,t) \;,
\end{equation}
with the ratio of $ \szerot,\tzerot$ matched to the ratio of meson masses in the respective channels cf. caption 
of \TAB\ref{tab:rSRparams}. 
In addition we impose $\szerot^{B_{s}}/\szerot^{B_{d,u}}\approx m_{B_s}^2/m_{B_{d,u}}^2$ and 
 $s_0^{\fb}/s_0^{f_{B^*[B_1]}} \approx m_B^2/m_{B^*[B_1]}^2$ 
to be satisfied reasonably well.
We turn to the dependency on a specific duality parameter $a$. 
It is found that in the $B$-meson cases a single set of SR parameters is sufficient to satisfy \eqref{eq:daughterSR} to within $\approx 2\%$ for the $a=\{1/2,1,2\}$ cases considered. For the $D$-mesons this no longer holds and a small modification to the SR parameters is made at each value of $a$.

\begin{table}[t]
  \centering
  \setlength\extrarowheight{-2.5pt}
    \resizebox{\columnwidth}{!}{
  \begin{tabular}{rr  |  c c  cc || c  cc}
            &           & $\mu_F^2\! =\!   \muals^2\! =\! \muCond^2[\GeV^2]$  & $\muUV[\GeV]$ & $\muM[\GeV]$       & $\mukin[\GeV]$ & $\muals[\GeV]$      & $\muCond[\GeV]$ \\  \hline  \rule{0pt}{1.2em}
    $\!\!B$ & Kin       & $m_B^2 - (m_b^{\textrm{pole}})^2$ & 4.78(1.0)          & $-  $                   & 1.0(4)            & 4.18(1.5)               & 3.0(1.0)       \\
    $B$     & $\MSbar$  & $m_B^2 - (m_b^{\textrm{pole}})^2$ & 4.78(1.0)          & $\Verr{4.18}{1.7}{1.2}$ &
    $ -$    & 4.18(1.5) & 3.0(1.0)       \\
    $D$     & $\MSbar$  & $m_D^2 - m_c(m_c)^2$              & 1.67(30)           & $\Verr{1.27}{1.5}{0.2}$ & $- $              & $\Verr{1.27}{1.0}{0.2}$ & 2.0(1.0)       \\
  \end{tabular}
  }
  \caption{\small Summary of the scales involved in the determination of the residues (\TAB\ref{tab:rValues})
  and coupling constants  (\TAB\ref{tab:gValues}) to the left of the  double separation line. 
  To the right we have the scale changes  used for the best determination of the decay constants 
  (\TAB\ref{tab:fBparams}).  The quantity $m_c(m_c)$, above,  is the $\MSbar$ mass at the scale $m_c$. 
  The uncertainty in $\mu_F$, for the $B$-meson ($D$-meson),  is chosen to be $\Delta\mu_F = \pm 1 \GeV$  $(\Delta\mu_F = ^{+1.0}_{-0.2} \GeV)$.  }
  \label{tab:scales}
\end{table}

 We consider it worthwhile to comment on the specific numerical values of the thresholds found. 
The expectation for a  single dispersive threshold $s_0$ is $(m_{B_i} + 2 m_\pi)^2 < s_0   <
(m_{B_i} +  m_\rho)^2$,  and lying closer to the top boundary.  
Inspecting   \TAB\ref{tab:rSRparams}, we note that  this is indeed the case for the single 
dispersion threshold $s_0^{f_B}$ but not for  the double dispersion threshold $\Sred$ \eqref{eq:Sred}.
 Whereas  $\Sred$ takes on a similar r\^ole to the single dispersive effective threshold, one must remember that it  contains additional  information on the excited vector meson   
 channel, cf. \eqref{eq:dualParams} and might further be a result of the peculiar analytic structure in  $(s,t)$ of the LC-OPE.\footnote{It is conceivable that if one were to adapt 
the daughter sum rule method to the extraction of $g_{DD*\pi}$ and $g_{BB^*\pi}$ in \cite{Khodjamirian:2020mlb},
one could even find better agreement with experiment and/or the lattice.}

Let us turn to the  correlation imposed on parameters based on  physical arguments.
Whilst the effective threshold for decay constant $s_0^{f_B}$ can be independently determined it would 
contradict the method if it were completely independent of the $\Sred$-threshold, since they are both associated with the same state.  A  $50\%$-correlation is adopted between the two.
 The vector versus tensor results are correlated since, by the (exact) equation of motion,  
their difference is equal to a derivative operator which is numerically (and to some extent parametrically) suppressed at low recoil. In order to remain consistent this implies a correlation of the 
effective thresholds, as argued in \cite{Hambrock:2013zya} and more systematically exploited in \cite{BSZ15}.\footnote{However, for the $\para$-direction the derivative term is not small and such 
a correlation does not make sense. See section 4.2  in  \cite{Janowski:2021yvz} for a more elaborate discussion  
in the context of the $\bar{B} \to \ga$ form factors.}
 The correlations
 \begin{alignat}{5}
\label{eq:corrR}
& \text{corr}({\Sred}^{V_{\perp}} ,{\Sred}^{T_{\perp}})|_{B}      = \frac{4}{5} \;, \quad   \text{corr}({\Sred}^{V_{\para}} ,{\Sred}^{T_{\para}})|_{B}   = \frac{1}{2} \;, \quad \text{corr}({\Sred}^{V} ,{\Sred}^{T})|_{D}  = \frac{2}{3} \;,
\end{alignat}
are imposed based on the contribution of the derivative operator to the equations of motion, which is 
$\approx 10\%$ in the $B\perp$- and $\approx 20\%$ in the $D\perp$-case. 
In the $B\|$-case the contribution is $\approx 40$-$45\%$.
\begin{table}
  \centering
  \setlength\extrarowheight{-2.5pt}
  \resizebox{\columnwidth}{!}{
  \begin{tabular}{l|c|c|c| | l|c|c}
    \multicolumn{7}{c}{\mbox{SR parameters $[\GeV^2]$}}\\ 
                                                    & $B_d$                 & $B_s$       & $B_u$                 &                            & $B_{d,u}$  & $B_s$\\
                                                  \hline
    $\{\Sred,\hat{M}^2\}^{r^V_{\perp},r^T_{\perp}}$ & 37.7,\;8.0\phantom{0} & 39.2,\;11.0 & 37.7,\;8.0\phantom{0} & $\{s_0,M^2\}^{\fb}$        & 34.3,\;5.6 & 35.5,\;6.4 \\
    $\{\Sred,\hat{M}^2\}^{r^V_{\parallel}}$         & 43.5,\;12.0           & 45.5,\;13.5 & 43.5,\;12.0           & $\{s_0,M^2\}^{\fbs,\fbsT}$ & 34.5,\;5.7 & 35.9,\;7.1\\
    $\{\Sred,\hat{M}^2\}^{r^T_{\parallel}}$         & 42.5,\;11.0           & 44.4,\;13.5 & 42.5,\;11.0           & $\{s_0,M^2\}^{\fbo,\fboT}$ & 38.9,\;6.0 & 40.6,\;8.6\\\hline
                                                    & $D_d$                 & $D_s$       & $D_u$                 &                            & $D_{d,u}$  & $D_s$\\
                                                  \hline
    $\{\Sreda{1/2},\hat{M}^2\}^{r^V_{\perp}}$       & 5.9,\;3.1             & 6.6,\;3.4   & 5.9,\;3.1             & $\{s_0,M^2\}^{\fd}$        & 5.7,\;1.9  & 6.3,\;2.2 \\
    $\{\Sreda{1/2},\hat{M}^2\}^{r^T_{\perp}}$       & 5.9,\;2.6             & 6.6,\;2.9   & 5.9,\;2.6             & $\{s_0,M^2\}^{\fds,\fdsT}$ & 6.1,\;1.9  & 6.9,\;2.6 \\\hline
    $\{\Sreda{1},\hat{M}^2\}^{r^V_{\perp}}$         & 6.0,\;2.9             & 6.7,\;3.2   & 6.0,\;2.9              \\
    $\{\Sreda{1},\hat{M}^2\}^{r^T_{\perp}}$         & 6.0,\;2.4             & 6.7,\;2.7   & 6.0,\;2.4              \\\cline{1-4}
    $\{\Sreda{2},\hat{M}^2\}^{r^V_{\perp}}$         & 5.7,\;2.9             & 6.5,\;3.4   & 5.7,\;2.9              \\
    $\{\Sreda{2},\hat{M}^2\}^{r^T_{\perp}}$         & 5.7,\;2.4             & 6.5,\;2.7   & 5.7,\;2.4              \\\cline{1-4}
  \end{tabular}
  }
  \caption{\small Summary of the SR parameters used in the determination of the residues for the triangular duality window $a=1$
  (cf. \SEC\ref{sec:gcomp} for comments). The additional threshold parameter is fixed via the ratio of the scalar and vector mesons. For the $B$-mesons $\tzerot/\szerot=m_{B^*}^2/m_B^2\approx 1.02$ and $\tzerot/\szerot=m_{B_1}^2/m_B^2\approx 1.18$ in the $\perp$ and $\|$ directions respectively. In the $D$-meson channels $\tzerot/\szerot=m_{D^*}^2/m_D^2\approx 1.15 $.  Note the difference between the values of $\Sred$ for the $\perp$- and $\|$-directions reflects the fact that $m^2_{B_1}/m^2_{B^*}\approx1.16$. We remind the reader that in the $B$-meson channels a single set of SR parameters is sufficient to satisfy the daughter SR \eqref{eq:daughterSR} to within $\approx 2\%$ for all three choices of the duality parameter $a$. For the $B$-meson ($D$-meson) processes we associate a uniform uncertainty to the threshold of $\pm 2.0\GeV^2$ ($\pm0.5\GeV^2$) and the Borel mass of  $\pm2.0\GeV^2$ ($\pm0.5\GeV^2$).}
  \label{tab:rSRparams}
\end{table}

\begin{table}[bp]
  \centering
  \setlength\extrarowheight{-5pt}
  \begin{tabular}{c | c |  l |r|r|r|r|r|r|r}
{\small twist} & {\small pa} & DA                                & $r_{\perp}^V(B_u)$ & $r_{\|}^V(B_d)$ & $r_{\perp}^T(B_s)$ & $r_{\|}^T(B_d)$ & $r_{\perp}^V(D_u)$ & $r_{\perp}^V(D_d)$ & $r_{\perp}^T(D_d)$ \\\hline
 
1              & $-$         & PT:$\,\ORD{(\al_s^0)}$            & $-0.116$           & $-0.014$        & $0.102$            & $0.033$         & $-0.177$           & $0.004$            & $0.014$ \\
1              & $-$         & PT:$\,\ORD{(\al_s)}$              & $-0.033$           & $0.003$         & $0.027$            & $-0.005$        & $-0.047$           & $-0.002$           & $<10^{-3}$ \\\hline
2              & 2           & $\phi_{\ga}(u)\,\ORD{(\al_s^0)}$  & $-0.136$           & $0.075$         & $0.063$            & $0.064$         & $-0.220$           & $0.110$            & $0.078$ \\
2              & 2           & $\phi_{\ga}(u)\,\ORD{(\al_s)}$    & $-0.028$           & $0.015$         & $0.015$            & $0.015$         & $-0.028$           & $0.014$            & $0.018$ \\\hline
3              & 2           & $\Psi_{(a)}(u)$                   & $0.011$            & $-$             & $0.016$            & $-0.001$        & $0.021$            & $-0.010$           & $-0.010$ \\
3              & 2           & $\Psi_{(v)}^{(1)}(u)$             & $-$                & $<10^{-3}$      & $0.001$            & $<10^{-3}$      & $-$                & $-$                & $0.005$ \\
3              & 3           & $\mathcal{A}(\underline{\alpha})$ & $-$                & $-$             & $<10^{-3}$         & $<10^{-3}$      & $-$                & $-$                & $-0.001$ \\
3              & 3           & $\mathcal{V}(\underline{\alpha})$ & $-$                & $-$             & $-0.002$           & $-0.003$        & $-$                & $-$                & $-0.014$ \\\hline
4              & 2           & $h_{\ga}^{(2)}(u)$                & $-$                & $-0.003$        & $<10^{-3}$         & $-0.002$        & $-$                & $-$                & $-0.003$ \\
4              & 2           & $\mathbb{A}(u)$                   & $0.017$            & $-0.05$         & $-0.003$           & $-0.003$        & $0.033$            & $-0.017$           & $-0.006$ \\
4              & 3           & $\sSum$                           & $<10^{-3}$         & $<10^{-3}$      & $<10^{-3}$         & $<10^{-3}$      & $-0.003$           & $0.001$            & $0.002$ \\
4              & 3           & $\tSum$                           & $-0.001$           & $<10^{-3}$      & $<10^{-3}$         & $<10^{-3}$      & $-0.007$           & $0.003$            & $<10^{-3}$ \\
4              & 3           & $S_{\ga}(\underline{\alpha}) $    & $-0.008$           & $<10^{-3}$      & $0.003$            & $0.003$         & $-0.033$           & $-0.033$           & $-0.026$ \\
4              & 3           & $T_4^{\ga}(\underline{\alpha})$   & $0.002$            & $<10^{-3}$      & $<10^{-3}$         & $<10^{-3}$      & $0.007$            & $0.007$            & $<10^{-3}$ \\
4              & $-$         & $Q_q\VEV{\bar q q}$               & $-$                & $0.002$         & $-$                & $-$             & $-$                & $-$                & $-$ \\
4              & $-$         & $Q_b\VEV{\bar q q}$               & $0.003$            & $-0.002$        & $0.002$            & $0.002$         & $-0.021$           & $-0.021$           & $-0.017$
\\\hline
              &             & Total$^*$                             & $-0.290$           & $0.070$         & $0.223$            & $0.102$         & $-0.475$           & $0.096$            & $0.081$
  \end{tabular}
  \caption{\small A breakdown of contributions according to twist, 
   ``pa" = number of partons and the specific DA. The definitions of the DAs can be found in \cite{Janowski:2021yvz}.  
   The asterisk in total is a reminder that it 
  does not include twist-$4$ contributions  not closing under the equations of motion  cf.~\cite{Janowski:2021yvz}.}
  \label{tab:twist_breakdown}
\end{table}

Another relevant aspect concerning the plethora of predictions is that not all channels are of equal quality.  This is highlighted by the two separate determinations of the residue, from the 
vector and tensor interpolating current.  Let us define the ratio $\UU_{B^{(i)}}=g^T_{BB^{(i)}\ga}/g^V_{BB^{(i)}\ga}$ which   ideally is close to  one. 
We find reassuringly good values for the $B^*$-case $( \UU_{B^{*0}} ,\UU_{B^{*}_s},\UU_{B^{*+}}) =  (0.99,0.98,0.98)$ , moderate deviations 
$( \UU_{D^{*+}} ,\UU_{D^{*}_s},\UU_{D^{*0}}) =  (0.81,0.82,0.91)$  for the $D^*$-case and significant deviations 
$( \UU_{B_{1}^0} ,\UU_{B_{1s}},\UU_{B_{1}^+}) =  (1.35,1.26,1.30)$ for the $B_1$-case as anticipated cf. 
footnote \ref{foot:D1}. For the $D^*$-case it is the accidental cancellation of the two 
charge contributions in perturbation theory, to be discussed further below, and the sensitivity to higher twist which gives rise to larger deviation from one. 
\begin{comment}
\begin{align}
r_{\perp}^V(D_{q=u,d})&\approx-(0.24\,Q_q+0.12\,Q_c)|_{\text{PT}}-0.35\,Q_q|_{\text{twist-2}}+\text{higher twist}\nonumber\\
r_{\perp}^T(D_{q=u,d})&\approx-(0.27\,Q_q+0.12\,Q_c)|_{\text{PT}}-0.31\,Q_q|_{\text{twist-2}}+\text{higher twist}\nonumber\\
r_{\perp}^V(D_s)&\approx-(0.34\,Q_s+0.15\,Q_c)|_{\text{PT}}-0.40\,Q_s|_{\text{twist-2}}+\text{higher twist}\nonumber\\
r_{\perp}^T(D_s)&\approx-(0.25\,Q_s+0.14\,Q_c)|_{\text{PT}}-0.32\,Q_s|_{\text{twist-2}}+\text{higher twist}
\end{align}
\end{comment}
For the $B_1$-case the concept of a well isolated resonance is not assured  and for the $D_1$ it simply does not hold cf. also footnote \ref{foot:D1}.
Therefore we do not quote any results for the $D_1$ whilst for the $B_1$ the results are deemed just 
marginally acceptable to present.

It is instructive to present a breakdown in terms of charges for comparison with other work 
and illustrate the, presumably accidental, cancellations in the charged $D$- and $B_1$-case
 which unfortunately  implies that these results are less reliable. For definiteness
we quote the breakdown for the couplings obtained from the \emph{vector} interpolating current \begin{alignat}{2}
\label{eq:breakdown}
&g_{D_qD_q^{*}\gamma}&\approx&-(1.05\,Q_q+0.51\,Q_c)|_{\text{PT}}-1.74\,Q_q|_{\text{twist-2}}+\text{ht} 
\;,\nonumber\\
%g^T_{D_qD_q^{*}\gamma}&\approx-(1.33\,Q_q+0.56\,Q_c)|_{\text{PT}}-1.49\,Q_q|_{\text{twist-2}}+\text{ht}\;,\nonumber\\
&g_{D_sD_s^{*}\gamma}&\approx&-(1.24\,Q_s+0.51\,Q_c)|_{\text{PT}}-1.62\,Q_s|_{\text{twist-2}}+\text{ht}\;,\nonumber\\
%g^T_{D_sD_s^{*}\gamma}&\approx-(1.36\,Q_s+0.54\,Q_c)|_{\text{PT}}-1.26\,Q_s|_{\text{twist-2}}+\text{ht}\;,
&g_{B_qB_q^{*}\gamma}&\approx&-(1.20\,Q_q+0.24\,Q_b)|_{\text{PT}}-1.20\,Q_q|_{\text{twist-2}}+\text{ht}\;,\nonumber\\
&g_{B_sB_s^{*}\gamma}&\approx&-(1.31\,Q_s+0.25\,Q_b)|_{\text{PT}}-1.10\,Q_s|_{\text{twist-2}}+\text{ht}\;,\nonumber\\
&g_{B_qB_{1q}\gamma}&\approx&+(0.12\,Q_q+0.02\,Q_b)|_{\text{PT}}-1.11\,Q_q|_{\text{twist-2}}+\text{ht}\;,\nonumber\\
&g_{B_sB_{1s}\gamma}&\approx&-(0.69\,Q_s-0.52\,Q_b)|_{\text{PT}}-1.03\,Q_s|_{\text{twist-2}}+\text{ht}\;,
\end{alignat}
\com{where $Q_i$ are the standard quark charges $Q_b = Q_d = Q_s = -\frac{1}{3}$ and $Q_{c} = Q_u = \frac{2}{3}$ and ``ht" stands for higher twist and $q=u,d$ .
The size of the higher twist can be inferred from \TAB\ref{tab:twist_breakdown}. The 
twist-$3$ contribution is up to $5\%$ in some cases and, as previously argued, most 
 twist-$4$ contributions have to be dropped since they are incomplete without 
 the  inclusion of $4$-particle DAs (cf. \SEC\ref{sec:gcomp} for further relevant remarks in this direction).}

\begin{figure}[t]
\centering
  \begin{minipage}{0.48\textwidth}
    \centering
    \includegraphics[width=1.\textwidth]{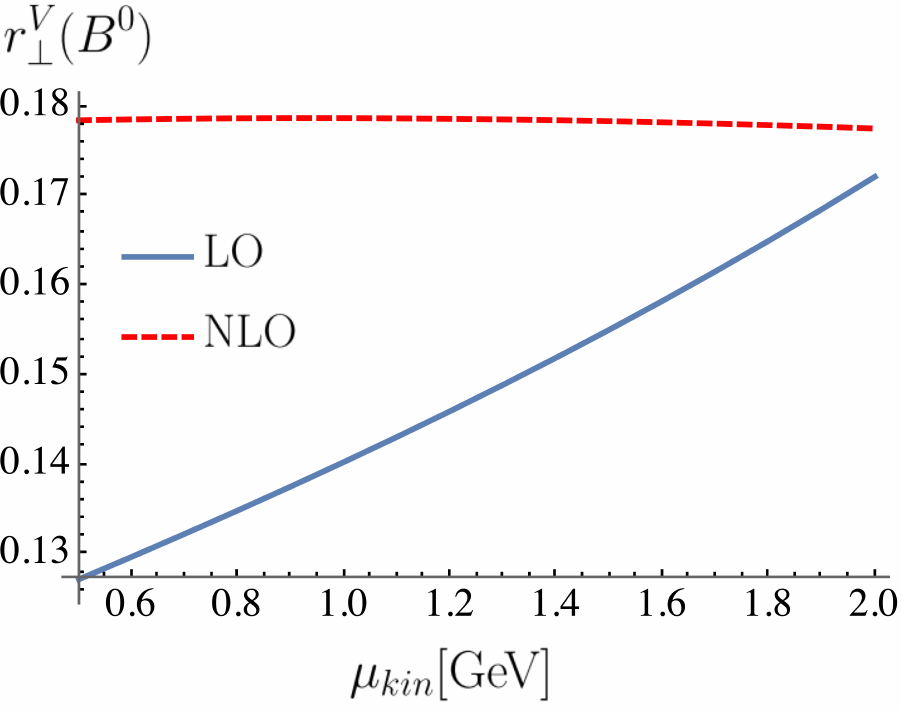}
  \end{minipage}\hfill
  \begin{minipage}{0.48\textwidth}
    \centering
    \includegraphics[width=1.\textwidth]{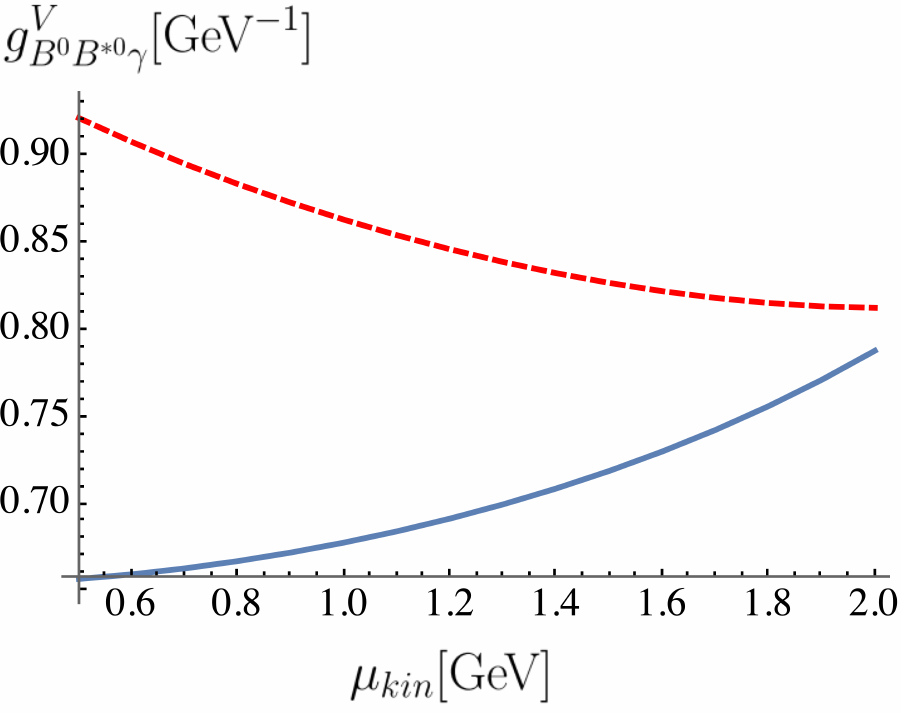}
  \end{minipage}
  \begin{minipage}{0.48\textwidth}
    \centering
    \includegraphics[width=1.\textwidth]{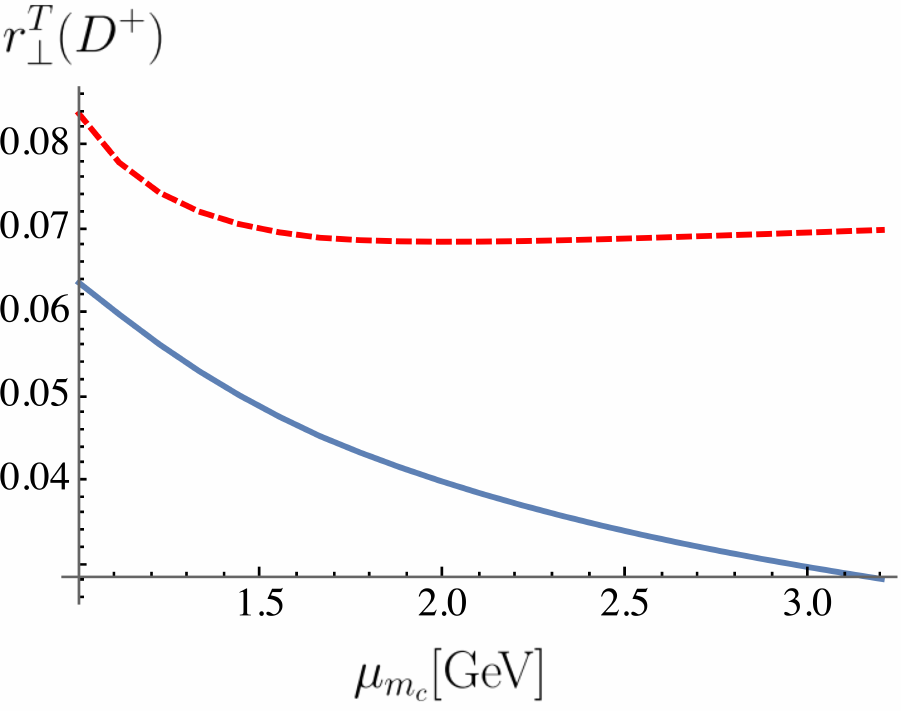}
  \end{minipage}\hfill
  \begin{minipage}{0.48\textwidth}
    \centering
    \includegraphics[width=1.\textwidth]{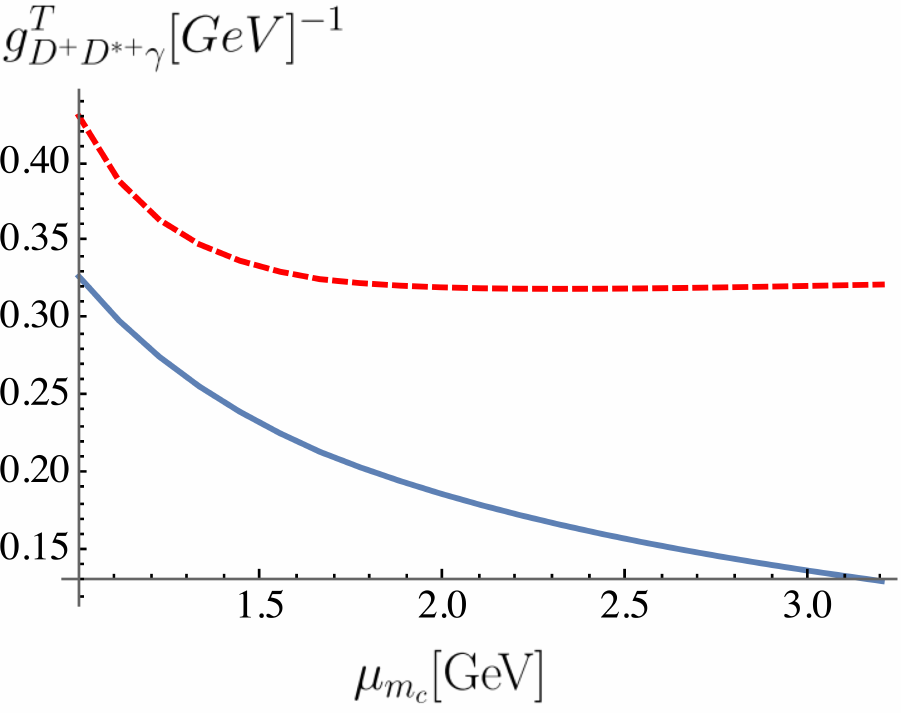}
  \end{minipage}
  \caption{\small Representative plots highlighting how the NLO result (dashed red) sees a reduction in the 
  mass scheme scale $\mukin(\mu_{\MSbar})$ as compared to LO  (solid blue). We note that in the 
  $B$-meson case the residue generally shows greater stability than the coupling which, in the above cases, inherits the (artificial) scale dependence of  $f_{B^*}$ which is less pronounced for $f_{D^*}$ as the plots show.} 
  \label{fig:scale_dep}
\end{figure}

We now proceed to discuss the numerical features of the $B$- and $D$-meson results 
 in turn. Beginning with the \emph{$B$-mesons}, for the values given in \TAB\ref{tab:rSRparams} we find that the daughter SRs \eqref{eq:daughterSR} are, in all cases, satisfied to within $\lesssim 2\%$. The continuum contributions range from $\lesssim25\%$ in the 
 $\perp$-modes to $\lesssim35\%$ in the $\|$-modes. In the $\perp$-modes the SR is dominated by the perturbative and twist-$2$ contributions which are approximately equal in size and 
 are of the  same sign. The remaining contributions make up $\approx10$-$20\%$ of the total. The story is repeated in the tensor $\|$-modes, however the situation in the vector 
 $\|$-modes is somewhat altered. Here unfortunate cancellations act to suppress the perturbative contribution and  the twist-$2$ sector is numerically dominant 
 providing $\ORD(80)\%$ of the total value. A breakdown of contributions according to twist is given in \TAB\ref{tab:twist_breakdown} for a representative selection of  residues. 
The $\ORD{(\al_s)}$ corrections are mass scheme dependent. 
In the kinetic mass scheme ($\mukin=1\GeV$) employed the NLO results are 
sizeable, providing a correction 
of $\approx 20$-$35\%$ and $\approx 20$-$25\%$ at twist-$1$ and -$2$, respectively cf. \TAB\ref{tab:twist_breakdown}. 
The benefit and necessity of an NLO computation is clearly visible 
in the scale variation plots shown in \FIG\ref{fig:scale_dep} as residual effects are then 
of $\ORD(\al_s^2)$.

The SR parameters for the \emph{$D$-mesons}  are determined subject to the same tests as outlined above. In all cases the continuum contribution remains below $30\%$ and in the neutral modes the daughter SR \eqref{eq:daughterSR} is satisfied to within $3\%$. In the charged mode the daughter SR shows poor convergence. 
Again, this is due to the presumably artificial smallness of 
 the perturbative contribution  due to  cancellation in $Q_c$ and $Q_u$ 
 (cf. \TAB\ref{tab:twist_breakdown} and \eqref{eq:breakdown}). In contrast to the $B$-mesons  we note that whilst the dominant contribution arises from the twist-$1$ or -$2$ sectors the twist-$3$ and -$4$ sectors are sizeable, in particular in the charged case. The $\ORD(\al_s)$ corrections to twist-$2$ range from $\approx 12\%$ of the LO result in the vector modes to $\approx20\%$ in the tensor modes. In the twist-$1$ sector the tensor modes and the neutral vector mode have radiative corrections ranging between $\approx 2$ - $ 30\%$. In the charged vector modes, however,  the corrections are  $>50\%$,  due  to the previously mentioned large charge cancellation at LO.
\begin{table}[t]
  \centering
  \setlength\extrarowheight{-3pt}
  \begin{tabular}{l | l | l | l | l | l | l | l}
                                       & $r_{\perp}^V(B_u)$    & $r_{\|}^V(B_d)$      & $r_{\perp}^T(B_s)$   & $r_{\|}^T(B_d)$      & $r_{\perp}^V(D_u)$   & $r_{\perp}^V(D_d)$   & $r_{\perp}^T(D_d)$  \\ \hline
                                          Value                                  & -0.300                & 0.076                & 0.224                & 0.104                & -0.473               & 0.095                & 0.073\\\hline
Error                                  & $\err{0.034}{0.033}$  & $\err{0.015}{0.016}$ & $\err{0.023}{0.024}$ & $\err{0.015}{0.015}$ & $\err{0.054}{0.053}$ & $\err{0.021}{0.021}$ & $\err{0.020}{0.020}$  \\   \hline
$\Delta \overline{s_0}$                & $\pm 0.021$           & $\pm 0.007$          & $\pm 0.017$          & $\pm 0.007$          & $\pm 0.033$          & $\pm 0.004$          & $\pm 0.004$\\
$\Delta \hat{M}^2$                     & $\err{0.005}{0.000}$  & $\err{0.001}{0.002}$ & $\err{0.000}{0.002}$ & $\err{0.000}{0.002}$ & $\err{0.004}{0.001}$ & $\err{0.001}{0.002}$ & $\err{0.004}{0.001}$\\
$\Delta M^2_{\fb}$                     & $\err{0.000}{0.003}$ & $\err{0.001}{0.000}$ & $\err{0.004}{0.000}$ & $\err{0.001}{0.000}$ & $\err{0.000}{0.004}$ & $\err{0.001}{0.000}$ & $\err{0.001}{0.000}$\\
$\Delta m_{b,c}$                       & $\pm 0.002$           & $\pm 0.004$          & $\pm 0.002$          & $\pm 0.002$          & $\pm 0.003$          & $\pm 0.003$          & $\pm 0.003$\\
$\Delta \tau $                         & $\pm 0.010$           & $\pm 0.006$          & $\pm 0.005$          & $\pm 0.005$          & $\pm 0.015$          & $\pm 0.008$          & $\pm 0.006$\\
$\Delta \vev{\bar{q}q} $               & $\pm 0.003$           & $\pm 0.001$          & $\pm 0.001$          & $\pm 0.001$          & $\pm 0.005$          & $\pm 0.004$          & $\pm 0.003$\\
$\Delta a_2$                           & $\pm 0.020$           & $\pm 0.011$          & $\pm 0.009$          & $\pm 0.010$          & $\pm 0.033$          & $\pm 0.016$          & $\pm 0.012$\\
$\Delta_{ t=3}$                        & $\pm 0.006$           & $<10^{-3}$           & $\pm 0.004$          & $\pm 0.003$          & $\pm 0.011$          & $\pm 0.006$          & $\pm 0.011$\\
%   $\Delta_{t=4}$                     & $\pm 0.004$           & $\pm 0.002$          & $\pm 0.001$          & $\pm 0.001$          & $\pm 0.010$          & $\pm 0.005$          & $\pm 0.002$\\
$\Delta \mukin$                        & $\pm 0.003$           & $\pm 0.003$          & $\pm 0.002$          & $\pm 0.004$          & \mc{$-$}             & \mc{$-$}             & \mcc{$-$}\\
$\Delta \mu_m$                         & \mc{$-$}              & \mc{$-$}             & \mc{$-$}             & \mc{$-$}             & $\err{0.008}{0.001}$ & $\err{0.004}{0.006}$ & $\err{0.004}{0.007}$ \\
$\Delta \mu_{\al_s}$                   & $\err{0.010}{0.003}$  & $\err{0.001}{0.004}$ & $\err{0.001}{0.003}$ & $\err{0.001}{0.001}$ & $\err{0.002}{0.003}$ & $<10^{-3}$        & $\err{0.002}{0.001}$\\
$\Delta \mu_{F}$                       & $\err{0.002}{0.000}$  & $\err{0.001}{0.004}$ & $\err{0.002}{0.007}$ & $\err{0.000}{0.001}$ & $\err{0.005}{0.002}$ & $\err{0.004}{0.001}$ & $\err{0.000}{0.001}$\\
$\Delta \muUV$                         & \mc{$-$}              & \mc{$-$}             & $\pm 0.002$          & $\pm 0.001$          & \mc{$-$}             & \mc{$-$}             & $\pm 0.001$\\
$\Delta \frac{\CondQQ{q}}{\CondQQ{s}}$ & \mc{$-$}              & \mc{$-$}             & $\pm 0.002$          & \mc{$-$}             & \mc{$-$}             & \mc{$-$}             & \mcc{$-$}\\
$\Delta \Sigma$                        & $\pm 0.001$           & $\pm 0.003$          & $\pm 0.006$          & $\pm 0.006$          & $\pm 0.013$          & $\pm 0.002$          & $\pm 0.006$
  \end{tabular}
  \caption{\small Breakdown of the main contributions to the uncertainty for a representative selection of residues. $\Delta \overline{s_0}$ includes the combined uncertainty, incorporating correlations, due varying all effective thresholds, cf. discussion above \eqref{eq:corr}. $\Delta_{t=3}$  contains the total uncertainty due to all twist-$3$ hadronic parameters $\{f_{3\gamma},\omega_{\gamma}^A,\omega_{\gamma}^V\}$. The uncertainty due to the choice of duality region is encapsulated in the quantity $\Delta \Sigma$ which represents the standard deviation of the $a=\{1/2,1,2\}$ evaluations. The total uncertainty, 
 which also includes smaller contributions such as the gluon condensate, is obtained by added 
 uncertainties in quadrature.}
  \label{tab:rErrors}
\end{table}

The uncertainty due to the input parameters is estimated by varying each parameter, within the given interval, in turn and adding each individual uncertainty in quadrature. To incorporate correlations between the various thresholds, discussed previously, we generate 300 samples of the thresholds according to a Gaussian distribution such that the mean corresponds to the central value of each threshold and the standard deviation reproduces the associated uncertainty. We then evaluate the desired quantity for each of these samples, taking the standard deviation of the resulting points to be the uncertainty due to threshold variation. Our predictions for the couplings are given as the mean value of the vector and tensor interpolating current determinations. We estimate the associated uncertainty as the standard deviation of the two evaluations. Moreover, the uncertainty associated with varying the duality window is taken to be the standard deviation of the $a=\{ 1/2,1,2\}$ determinations (cf. \TAB\ref{tab:gValuesA}). This provides a small contribution in the $B$-meson cases, but notably a more significant contribution in the $D_s$ mode. Adding in quadrature the uncertainty from all sources, we obtain the total uncertainty as quoted in \TAB\ref{tab:gValues}.
\begin{table}[btp]
  \centering
  \small
  \setlength\extrarowheight{-4.5pt}
  \begin{tabular}{| l| r | r | r | r | r | r | r | r | r | r |}\hline
          & $g_{B_dB_d^{*}\ga}$ & $g_{B_sB_s^{*}\ga}$ & $g_{B_uB_u^{*}\ga}$ & $g_{B_dB_{1d}\ga}$ & $g_{B_sB_{1s}\ga}$ & $g_{B_uB_{1u}\ga}$ & $g_{D_dD_d^{*}\ga}$ & $g_{D_sD_s^{*}\ga}$ & $g_{D_uD_u^{*}\ga}$ \\\hline
  $a=1/2$ & $0.86$              & $0.96$              & $-1.43$             & $0.35$             & $0.42$             & $-0.70$            & $0.38$             & $0.69$              & $-2.03$\\
  $a=1$   & $0.86$              & $0.95$              & $-1.44$             & $0.37$             & $0.44$             & $-0.72$            & $0.40$             & $0.60$              & $-2.11$\\
  $a=2$   & $0.86$              & $0.94$              & $-1.43$             & $0.40$             & $0.48$             & $-0.74$            & $0.40$             & $0.52$              & $-2.03$\\\hline
  \end{tabular}

  \caption{\small Values of the coupling for different values of the duality parameter $a$ 
  (cf. \SEC\ref{sec:duality} and \eqref{eq:t0}).
  The majority of couplings show little dependence on the duality parameter $a$. Notable exceptions are the neutral $B_1$-couplings and the charged $D^*$-couplings.  
  This does not come as a surprise as precisely those are  plagued by, unfortunate, cancellations in the perturbative contribution of the $Q_{b,c}$- and $Q_q$-parts (cf. end of \SEC\ref{sec:gcomp} for comments).}
  \label{tab:gValuesA}
\end{table}

The final values for the residues and the couplings are shown in \TABs\ref{tab:rValues}
and \ref{tab:gValues} respectively. The value of the coupling presented in the table is the average of the vector and tensor determinations.

\begin{table}[]
  \centering
  \setlength\extrarowheight{-3pt}
  \begin{tabular}{| c| l |  r | r | r | r |}
  \hline \hline
    &       & \mc{$r_{\perp}^V$}                      & \mc{$r_{\perp}^T$}            & \mc{$r_{\parallel}^V$}        & \mc{$r_{\parallel}^T$}\\\hline
  \multirow{3}{2.7cm}{$\mukin=1.0\GeV$}
    & $B_d$ & $\Verr{\phantom{-}0.179}{0.019}{0.019}$ & $\Verr{0.171}{0.018}{0.020}$  & $\Verr{0.076}{0.015}{0.016}$  & $\Verr{0.104}{0.015}{0.015}$\\
    & $B_s$ & $\Verr{\phantom{-}0.235}{0.024}{0.025}$ & $\Verr{0.224}{0.023}{0.024}$  & $\Verr{0.114}{0.016}{0.018}$  & $\Verr{0.146}{0.017}{0.017}$\\
    & $B_u$ & $\Verr{-0.300}{0.034}{0.033}$           & $\Verr{-0.284}{0.033}{0.031}$ & $\Verr{-0.159}{0.031}{0.029}$ & $\Verr{-0.199}{0.028}{0.028}$\\
     \hline \hline
  \multirow{3}{2.7cm}{$\mu_{\MSbar}=m_c(m_c)$}
    & $D_d$ & $\Verr{0.095}{0.021}{0.021}$            & $\Verr{0.073}{0.020}{0.020}$  & \mc{$-$}                      & \mc{$-$}\\
    & $D_s$ & $\Verr{0.172}{0.033}{0.029}$            & $\Verr{0.140}{0.048}{0.046}$  & \mc{$-$}                      & \mc{$-$}\\
    & $D_u$ & $\Verr{-0.473}{0.054}{0.053}$           & $\Verr{-0.412}{0.049}{0.049}$ & \mc{$-$}                      & \mc{$-$}\\ \hline
  \end{tabular}
  \caption{\small The residues \eqref{eq:r}, related to form factors, 
  for the $B$- and $D$-mesons. The former are determined in the kinetic scheme and the latter in the $\MSbar$ scheme.}
  \label{tab:rValues}
\end{table}

\subsubsection{Comparison with  Literature and Experiment}
\label{sec:gcomp}

It is of interest  to compare to the existing literature and experiment. 
The values of the couplings obtained in this work, which constitute the mean value of the tensor and vector determinations, along with determinations from other computations as well as 
  experiment  are collected 
in \TAB\ref{tab:gValues}. Unfortunately only two of the six couplings can be inferred
from experiment as the widths of the vector 
mesons are too  often unknown.\footnote{The situation is different to the  $g_{BB^*\pi}$ couplings as 
there the $B^*\to  B \pi$ decay is kinematically forbidden and thus 
it seems unfortunate that the $B^*(B_1) \to B \ga$ 
transitions are not known because of unknown total widths.}   
Moreover, in this section we use $(B_u,B_d,D_d,D_u)  \to (B^+,B^0,D^+,D^0)$  
which is the notation often used in experiment.

\begin{table}[]
  \centering
  \small
  \setlength\extrarowheight{-4pt}
  \begin{tabular}{| l| l | l | l || l | l | l |}
  \hline \hline
 \textrm{units}        [$\GeV^{-1}$]                                       & \mc{$g_{B^0B^{*0}\ga}$}                    & \mc{$g_{B_sB_s^{*}\ga}$}          & \mct{$g_{B^+B^{*+}\ga}$}          & \mc{$g_{D^+D^{*+}\ga}$}               & \mc{$g_{D_sD_s^{*}\ga}$}                               & \mc{$g_{D^0D^{*0}\ga}$}\\ \hline
 \rowcolor{gray!10}
$\!\!$This work                                                           & $\textrm{\phantom{-}}0.86^{+0.15}_{-0.15}$ & \phantom{-}$0.95^{+0.15}_{-0.16}$ & -$1.44^{+0.27}_{-0.26}$           & \phantom{-}$0.40^{+0.12}_{-0.13}$     & \phantom{-}$0.60^{+0.19}_{-0.18}$                      & -$2.11^{+0.35}_{-0.34}$\\
LCSR (NLL) \cite{Li:2020rcg}                                              & -$0.91^{+0.12}_{-0.13}$                    & -$0.74^{+0.09}_{-0.10}$           & \phantom{-}$1.44^{+0.22}_{-0.20}$ & -$0.15^{+0.11}_{-0.10}$               & -$0.079^{+0.086}_{-0.078}$                             & \phantom{-}$1.48^{+0.29}_{-0.27}$\\
HH$\chi$PT \cite{Amundson:1992yp}                                         & -$1.01^{+0.05}_{-0.05}$                    & -$0.70^{+0.06}_{-0.06}$           & \phantom{-}$1.45^{+0.11}_{-0.11}$ & -$0.27^{+0.05}_{-0.05}$               & \phantom{-}$0.041^{+0.056}_{-0.056}$                   & \phantom{-}$2.19^{+0.11}_{-0.11}$  \\
VMD $\!+\!$ HQET \cite{Colangelo:1993zq}                                  & -$0.58^{+0.12}_{-0.10}$                    & \mc{$-$}                          & \phantom{-}$0.99^{+0.19}_{-0.13}$ & -$0.29^{+0.19}_{-0.11}$               & -$0.19^{+0.19}_{-0.08}$                                & \phantom{-}$1.60^{+0.35}_{-0.45}$\\
CQM $\!+\!$ HQET \cite{Cheung:2014cka}                                    & -$0.82^{+0.06}_{-0.05}$                    & \mc{$-$}                          & \phantom{-}$1.45^{+0.11}_{-0.12}$ & -$0.38^{+0.05}_{-0.04}$               & \mc{$-$}                                               & \phantom{-}$1.91^{+0.09}_{-0.09}$   \\
RQM \cite{Goity:2000dk}                                                   & -$0.93^{+0.05}_{-0.05}$                    & \phantom{-}$0.65^{+0.03}_{-0.03}$ & \phantom{-}$1.66^{+0.11}_{-0.11}$ & -$0.44^{+0.06}_{-0.06}$               & -$0.19^{+0.03}_{-0.03}$                                & \phantom{-}$2.15^{+0.11}_{-0.11}$ \\
Lattice \cite{Becirevic:2009xp}$^{*}$, \cite{Donald:2013sra}$^{\ddagger}$ & \mc{$-$}                                   & \mc{$-$}                          & \mct{$-$}                         & -$0.20^{+0.30}_{-0.30}\phantom{}^{*}$ & \phantom{-}$0.11^{+0.02}_{-0.02}\phantom{}^{\ddagger}$ & \phantom{-}$2.00^{+0.60}_{-0.60}\phantom{}^{*}$ \\
Experiment   \cite{PDG}                                                              & \mc{$-$}                                   & \mc{$-$}                          & \mct{$-$}                         & \phantom{-}$0.47^{+0.07}_{-0.07}$     & \mc{$-$}                                               & \phantom{-}$1.77^{+0.16}_{-0.16}\phantom{}^\dagger$\\ \hline \hline
\textrm{units}        [$\GeV^{-1}$]                                       & \mc{$g_{B^0B^{0}_1\ga}$}                   & \mc{$g_{B_sB_{1s}\ga}$}           & \mct{$g_{B^+B^{+}_1\ga}$}         &                                       &                                                        & \\ \hline
 \rowcolor{gray!10}
$\!\! $ This work                                                         & \phantom{-}$0.37^{+0.12}_{-0.12}$          & \phantom{-}$0.44^{+0.12}_{-0.12}$ & -$0.72^{+0.21}_{-0.20}$ &                                       &                                                        & \\ \hline

  \end{tabular}
  \caption{\small $^\dagger$The experimental value of 
$g_{D^0D^{*0}\ga}$ requires the use of isospin 
   symmetry to deduce the width 
  of the $D^{*0}$ cf. the main text in \SEC\ref{sec:gcomp}. 
Here we take  $(B_u,B_d,D_d,D_u)  \to (B^+,B^0,D^+,D^0)$  to conform to PDG.
Note that the sign cannot be determined from experiment and 
that for the theory results it is convention dependent
  cf. \eqref{eq:Leffs}.  We further note that $g_{B_sB_s^*\ga} = \frac{f_{B_s}}{f_{B_s^*}} \frac{m_{B_s^*}}{m_{B_s}} |\mu| \approx 1.18 \GeV^{-1}$ with $|\mu| =  1.13   \GeV^{-1} $ from  \cite{Aditya:2012im}, 
  deduced from $D^*$-decays and subject to $1/m_c$ corrections, is rather close to our value
  given the difference in methods.  
  Whereas Ref. \cite{Aliev:1995zlh}  established that the $D$-couplings can be determined 
  from LCSR we do not include the LO results presented therein as the input is outdated
  and a numerical comparison seems of limited use.}
  \label{tab:gValues}
\end{table}

With regards to the two experimental values, we update the analysis in  \cite{Becirevic:2009xp}
and make some further comments. We first turn to the $D^{*+}$, for which the 
width $\Ga( D^{*+} ) = 83.4(18)\keV $ and branching fraction $B( D^{*+} \to D^+ \ga) = 0.016(4)$ 
are known \cite{PDG}, and with \eqref{eq:widths}
give $|g_{D^+D^{*+}\ga}|= 0.47(7)\GeV^{-1}$ instead of the previous $0.50(8)\GeV^{-1}$ 
in  \cite{Becirevic:2009xp}.
For  $D^{*0}$ the width is unknown and one needs to rely on isospin to infer it  \cite{Becirevic:2009xp}.  
First we deduce   $g_c$, related to  the $D^{*+}D\pi$-coupling, as  $g_c = 0.57(7)$, which is down from 
$0.61(7)$ in  \cite{Becirevic:2009xp}.  Considering all decay channels one then obtains 
 $\Ga(D^{*0}) = 56.5(14.0)\keV$, down from $68(17)\keV$ in  \cite{Becirevic:2009xp}.
Using the branching fraction $B(D^{*0} \to D^0 \ga) = 0.353(9)\keV$, down from 
$0.381(29)\keV$ we get $|g_{D^0D^{*0}\ga}|= 1.77(16)\GeV^{-1}$, down from $2.02(26)\GeV^{-1}$ in  \cite{Becirevic:2009xp}. 

Our result $g_{D^0D^{*0}\ga} = 2.11^{+0.35}_{-0.34}\GeV^{-1}$ is compatible with experiment 
and so are the results of the other method. 
Our value for $g_{D^+D^{*+}\ga} = 0.40^{+0.12}_{-0.13}\GeV^{-1}$ is again compatible with 
the new experimental value $|g_{D^+D^{*+}\ga}|= 0.47(7)\GeV^{-1}$.  
Differences between this work and the LCSR computation \cite{Li:2020rcg} are noticeable and can be at least partially accounted for by computational differences. 
Firstly, we have computed twist-$1$ $\ORD(\al_s)$-corrections whereas they did not.
Secondly, we include linear quark mass correction at LO and in the magnetic vacuum susceptibility 
$\chi_q$ (cf. section 3.2.1 in \cite{Janowski:2021yvz}). Third and most importantly, 
we drop twist-$4$ corrections other than the $Q_b\CondQQ{q}$ condensate cf. previous section.
 For the $B$-meson case the twist-$4$ corrections are not large and
the impact is small and the differences can be attributed to the first two cases. For the $D$-mesons 
higher twist corrections are more important per se, as they are less convergent. In addition, for the 
charged case perturbation theory is presumably artificially suppressed which makes these 
results less reliable in general.   Another important aspect is that in the charged case the 
inclusion of $S_\ga$ and $T_{4\ga}$ is definitely incomplete as the photon can connect to the external states. 
A sign of this is that in the neutral case the twist-$4$ contributions cancel, whereas in the charged 
case they are additive cf.  \TAB\ref{tab:twist_breakdown}.

The lattice determination 
$g_{D_sD^{*}_s\ga} |_{\mbox{\cite{Donald:2013sra}} }= 0.11(2)$ is approximately three standard deviations 
 lower than our value  $g_{D_s D^{*}_s\ga}= 0.60(19)$. 
This is where the breakdown \eqref{eq:breakdown} is useful. 
We find $g_{D_s D^{*}_s\ga} \approx -0.6 \,Q_c  - 3.0\, Q_s$ and from \FIG3 in \cite{Donald:2013sra} one deduces $g_{D_s D^{*}_s\ga} |_{\mbox{\cite{Donald:2013sra}} } \approx -0.66 Q_c  - 1.65 Q_s$.  
Whilst it is noted that in both cases the charm and strange quark contribution largely  cancel
each other, the effect is more pronounced for the lattice result.   The charm contribution is rather close and 
the deviation is in the strange quark part with almost a factor 2 difference, which seems large but not as large 
as the initial number would suggest.  It is instructive to investigate the $D^{*+} \to D^+ \ga $ 
case as  by $D$-spin symmetry\footnote{The exchange of $d \leftrightarrow s$, which is still a good approximate symmetry under QED.}, one would roughly expect a $20$-$30\%$-deviation. 
For our computation this is indeed the case $g_{D^+ D^{*+} \ga} \approx -0.65 \, Q_c  - 2.5 \, Q_d \approx 0.40(13)$, which does agree reasonably well with experiment $g_{D^+ D^{*+} \ga} = 0.47(7)$ (cf. \TAB\ref{tab:gValues}). 
Concerning the question of $D$-spin breaking, some further  guidance can be obtained from 
the lattice evaluation of the $D_{d,s} \to \ga$ form factors  \cite{Desiderio:2020oej}.  
The fits to a linear and an extended pole model are in agreement with $20$-$30\%$ $D$-spin breaking 
close to the kinematic endpoint.   
If the same level of $D$-spin breaking were valid at the $m_{D^*}^2$-pole, 
which some past experience suggests, then  
 $g_{D^+ D^{+*} \ga}$ and $g_{D_s D^*_s \ga}$ should not deviate considerably 
more than $20$-$30\%$ from each other.  
If the former is true  then this gives rise to a tension 
between  the experimental $g_{D^+ D^{*+} \ga} = 0.47(7)$  and the lattice determination of 
$g_{D_sD^{*}_s\ga} |_{\mbox{\cite{Donald:2013sra}} }= 0.11(2)$.  In conclusion it remains somewhat unclear what the resolution of this puzzle is. Whereas the sum rule results seems consistent,
  we wish to emphasise that, in exactly these modes, the sum rules are not the best of their kind for various reasons.  It may well be that the level of cancellations between the strange and the charm charge contributions 
  are so severe that past experience is overthrown. 
 It would be helpful to have further lattice determinations of these couplings and in particular 
 a more precise one for $g_{D^+ D^{+*} \ga}$.

\section{The \texorpdfstring{$f_H,f_{H^*},f_{H_1},f_{H^*}^T$}{} and \texorpdfstring{$f_{H_1}^T$}{} Decay Constants  from QCD Sum Rules}
\label{sec:decaySR}

The main reason for computing the decay constants is that 
to the best of our knowledge  $\{f_{H^*}^T, f_{H_1}^T\}$, required for 
the relation between the couplings and the (form factor) residues \eqref{eq:r},   
have not  been subjected to a QCD SR evaluation and are thus new.
The quantities $f_B$ and $f_{B^*}$ have previously been computed \cite{Jamin:2001fw,Gelhausen:2013wia,Wang:2015mxa}
to NLO with even partial NNLO results. 
We recompute these SRs and find agreement with the analytic expressions of the first two references.\footnote{\label{foot:compare} A direct comparison with \cite{Jamin:2001fw,Gelhausen:2013wia} can be made by taking the limit $s_0\to\infty$ in the results 
in \APP \ref{app:decayConsts}, as we provide the correlation functions after taking the  Borel transform with continuum subtraction.}  
In the work \cite{Wang:2015mxa} the $\ORD(\al_s) \vev{\bar qq}$  corrections were computed independently and we do disagree with some the expression e.g. the incomplete Gamma function. Compare  equation (21) \cite{Wang:2015mxa} versus \eqref{eq:B6} and equation (59) in 
\cite{Gelhausen:2013wia}.

\subsection{The Computation}\label{sec:fBcomp}

The starting points for the computations are the ``diagonal" correlation functions
\begin{alignat*}{3}
&\Ga^{f_B}(p)\;&=&\;i\int_x e^{i p\Cdot x}\matel{0}{T\{J_B(x)J_B^{\dagger}(0)\}}{0}\;&=&\;\Ga_{f_B}(p^2)\;,\nonumber\\
&\Ga^{\fbs}_{\al\be}(p)\;&=&\;i\int_x e^{i p\Cdot x}\matel{0}{T\{J_{\al}(x)J^{\dagger}_{\be}(0)\}}{0}\;&=&\;V_{\al\be}\Ga_{\fbs}(p^2)+\tilde{V}_{\al\be}\tilde{\Ga}_{\fbs}(p^2)\;,\nonumber\\
&\Ga^{\fbsT}_{\al\be\ga\de}(p)\;&=&\;i\int_x e^{i p\Cdot x}\matel{0}{T\{J^T_{\al\be}(x)J^{T\dagger}_{\ga\de}(0)\}}{0}\;&=&\;T_{[\al\be][\ga\de]}\Ga_{\fbsT}(p^2)+\tilde{T}_{\al\be\ga\de}\tilde{\Ga}_{\fbsT}(p^2) \;,
\end{alignat*}
where we have taken $H = B$ for concreteness again.
Above  $J_{\al}=\bar{q}\ga_{\al}b$, $J^T_{\al\be}=\bar{q}\sigma_{\al\be}b$ and the previously encountered $J_B$ is given in  \eqref{eq:fBjB}. 
The Lorentz structures are 
\begin{equation*}
V_{\al\be}= \frac{p_{\al}p_{\be}}{p^2}-g_{\al\be} \;, \quad \tilde{V}_{\al\be}=p_{\al}p_{\be} \;, 
\quad T_{\al\be\ga\de}=- g_{\al\ga}\frac{p_{\ga}p_{\de}}{p^2} \;, \quad 
\tilde{T}_{\al\be\ga\de}=\tensor{\varepsilon}{_{\al}_{\be}^{\bar{\al}}^{\bar{\be}}}\,\tensor{\varepsilon}{_{\ga}_{\de}^{\bar{\ga}}^{\bar{\de}}}T_{\bar{\al}\bar{\be}\bar{\ga}\bar{\de}} \;.
\end{equation*}
 The Lorentz invariant functions are related to the hadronic quantities as follows 
 \begin{alignat}{1}
 \label{eq:fBres}
&  \Ga_{\fb} = \frac{m_B^4 f_B^2}{m_B^2 - p^2}   +    \dots \;,  \qquad 
 \Ga_{\fbs} = \frac{m_{B^*}^2 f_{B^*}^2}{m_{B^*}^2 - p^2}   +    \dots \;, \qquad
 \Ga_{\fbsT} = \frac{m_{B^*}^2 (\fbsT)^2}{m_{B^*}^2 - p^2}   +    \dots  \;,
  \end{alignat}
and  the remaining structure $\tilde{\Ga}_{\fbs}(p^2)$ is related to $\Ga_{\fb}(p^2)$ up to contact terms 
by the equation of motion. 
The correlation functions  for the 
$\{ f_{B_1}, f_{B_1}^T\} $ decay constants follow with  rules for the insertion of the $\ga_5$ into the currents cf. \eqref{eq:doubling}  and $B^* \to B_1$ in \eqref{eq:fBres} following the ideas in \cite{Gratrex:2018gmm}.

The generic SR is parametrised by
\begin{eqnarray}
\label{eq:fBx}
f_{\Bx}^2 = \frac{e^{\frac{m_{\Bx}^2}{M_{f_{\Bx}}^2}}}{\omega_{\Bx}} \Big( & & \int_{m_b^2}^{s^{f_{\Bx}}_0} ds\, e^{-\frac{s}{M_{f_{\Bx}}^2}}\,\rho_{f_{\Bx}}(s)  +  \nonumber \\[0.1cm]
& & 
e^{\frac{- m_b^2}{M_{f_{\Bx}}^2}} \left[ c_{\CondQQ{q}}^{f_{\Bx} }m_b \CondQQ{q}+c_{\CondG}^{f_{\Bx}}\CondG+ \frac{m_b}{M_{f_{\Bx}}^2 }  c_{\CondqGq{q}}^{f_{\Bx}}\CondqGq{q} \right] \Big)\;,
\end{eqnarray}
where $f_{\Bx} $ stands for any  $\{ f_B, f_{B^*}, f_{B^*}^T,
f_{B_1}, f_{B_1}^T\} $, $\omega_{B}=m_B^4/(m_b+m_q)^2$ and $\omega_{\Bx}=m_{\Bx}^2$ otherwise. 
The local OPE is performed up to  $\dim\le 5$ including $\ORD{(\al_s)}$ corrections to both the perturbative and quark condensate contributions. 
Four quark condensates ($d=6$) give contributions at the sub per mille level  and are omitted. 
We have checked that all the scale dependences, due to  NLO computations, are correct. 
This  includes the cancellation of the condensate scale, denoted by $\muCond$, up to $\ORD(\al_s^2)$
as well as  the anomalous scaling of the scalar and transverse decay constants 
\eqref{eq:gaT}.  Explicit results are given in  \APP\ref{app:decayConsts}.

\begin{table}[bp]
  \centering
  \setlength\extrarowheight{-4.0pt}
  \resizebox{\columnwidth}{!}{
   \begin{tabular}{| l | ll | ll | ll |}  \hline\hline   \rule{0pt}{1.4em} 
    $f_{\Bx}\backslash  \Bx(\MeV)$                          & $B(5280)$            & $B_s(5367)$              & $B^*(5325)$          & $B^*_s(5415)$        & $B_1(5726)$          & $B_{1s}(5829)$    \\ \hline
     lattice \cite{Aoki:2019cca,Lubicz:2017asp}            & 190.0(1.3)           & 230.3(1.3)               & 186.4(7.1)           & 223.1(5.6)           & \mcc{$-$}            & \mc{$-$} \\ 
    experiment \cite{PDG}   &  188(17)(18)   & \mc{$-$}    & \mcc{$-$}    & \mc{$-$}    & \mcc{$-$}    & \mc{$-$}    \\
    SR  \cite{Gelhausen:2013wia}           & $207^{+17}_{-9}$     & $242^{+17}_{-12}$        & $210^{+10}_{-12}$    & $251^{+14}_{-16}$    & $335^{+18}_{-18}$   \cite{Wang:2015mxa}          & $348^{+18}_{-18}$   \cite{Wang:2015mxa}     \\
    SR  \cite{Lucha:2010ea}                & $193.4(16.6)$        & $232.5(21.0)$            & \mcc{$-$}            & \mc{$-$}             & \mcc{$-$}            & \mc{$-$}   \\[0.1cm] 
     \rowcolor{gray!10}
    this work                          & $\Verr{192}{20}{19}$ & $\Verr{225}{21}{20}$     & $\Verr{209}{23}{22}$ & $\Verr{245}{24}{23}$ & $\Verr{247}{31}{29}$ & $\Verr{305}{27}{26}$  \\
    $\delta_\text{PT},\;\delta_\CondQQ{q}$ & 0.18, -0.03          & 0.20, -0.02              & 0.10, -0.08          & 0.13, -0.07          & 0.11, -0.09          & 0.14, -0.05           \\
    $s^{f_{\Bx}}_0$, $M^2_{f_{\Bx}}$       & 34.4, 5.7            & 35.6, 6.6                & 34.9, 6.2            & 36.2, 6.9            & 38.1, 5.7            & 40.9, 8.1    \\  \hline  \hline
    $f^{T}_{\Bx}\backslash  \Bx(\MeV)$                          &                      &                          & $B^*(5325)^T$        & $B^*_s(5415)^T$      & $B_1(5726)^T$        & $B_{1s}(5829)^T$  \\ \hline   \rowcolor{gray!10}
    this work                        &                      &                          & $\Verr{200}{21}{20}$ & $\Verr{236}{22}{21}$ & $\Verr{230}{29}{28}$ & $\Verr{285}{25}{24}$\\
    $\delta_\text{PT},\;\delta_\CondQQ{q}$ &                      &                          & 0.11, -0.08          & 0.14, -0.06          & 0.11, -0.09          & 0.14, -0.05\\
    $s_0^{f_{\Bx}}$, $M^2_{f_{\Bx}}$       &                      &                          & 34.9, 6.2            & 36.3, 7.4            & 38.1, 5.7            & 40.9, 8.6\\ \hline  \hline   \rule{0pt}{1.4em}
    $f^{(T)}_{D_{(i)}}\backslash  D_{(i)}(\MeV)$                          & $D(1865)$            & $D_s(1968)$              & $D^*(2007)$          & $D^*_s(2112)$        & $D^*(2007)^T$        & $D^*_s(2112)^T$  \\  \hline  
    lattice \cite{Aoki:2019cca,Lubicz:2017asp}            & 209.0(2.4)           & 248.0(1.6)               & 223.5(8.7)           & 268.8(6.5)           & \mcc{$-$}            & \mc{$-$}     \\
     experiment \cite{PDG}   &  203.7(47)(6)   & 257.8(41)(1)    & \mcc{$-$}    & \mc{$-$}    & \mcc{$-$}    & \mc{$-$}    \\
    SR  \cite{Gelhausen:2013wia}           & $201^{+12}_{-13}$    & $238^{+13}_{-23}$        & $242^{+20}_{-12}$    & $293^{+19}_{-14}$    & \mcc{$-$}            & \mc{$-$}     \\
    SR  \cite{Lucha:2010ea}                & $206.2(12.4)$        & $245.3(20.2)$            & \mcc{$-$}            & \mc{$-$}             & \mcc{$-$}            & \mc{$-$} \\  \rowcolor{gray!10}
       this work                    & $\Verr{190}{15}{15}$ & $\Verr{226}{17}{17}$     & $\Verr{227}{18}{17}$ & $\Verr{279}{19}{19}$ & $\Verr{202}{16}{16}$ & $\Verr{256}{16}{17}$  \\
    $\delta_\text{PT},\;\delta_\CondQQ{q}$ & 0.24, 0.02           & 0.28, 0.03               & 0.05, -0.15          & 0.11, -0.11          & 0.07, -0.14          & 0.14, -0.10 \\
    $s_0^{f_{\Bx}}$, $M^2_{f_{\Bx}}$       & 5.7, 1.9             & 6.3, 2.2                 & 5.9, 2.0             & 6.8, 2.7             & 5.8, 2.2             & 6.9, 3.0 \\ \hline
  \end{tabular}
  }
  \caption{\small    QCD SR  results for the decay constants, in units of $\MeV$, 
 with the exception of  the Borel parameter and  the threshold which are given in $\GeV^2$-units.
  The  kinetic ($\mukin\!=\!1\GeV$) and $\MSbar$ ($\mu_m\!=\!m_c(m_c)$) schemes are employed for the $B$- and $D$-meson case respectively. 
   Input values are given in \TAB\ref{tab:inputParams}. For the $B(D)$-meson 
  SR  a uniform uncertainty   $\Delta s_0\!=\!\pm 1.5\GeV^2$ ($\Delta s_0\!=\!\pm0.5\GeV^2$),  $\Delta M^2\!=\!\pm1.5\GeV^2$ ($\Delta M^2\!=\!\pm0.5\GeV^2$) is applied to the threshold and the Borel parameter. 
   The slightly smaller uncertainty assigned to the decay constant SR parameters versus those of the residues reflects the fact that the daughter SR is satisfied to within $\approx0.5\%$ in the former but only to within $\approx2\%$ in the latter. The relative size of the radiative corrections are denoted by $\delta X$ such that $f_{\Bx}|_{X_{\text{NLO}}}\!=\!f_{\Bx}|_{X_{\text{LO}}}(1\!+\!\delta_X)$, with $X=\{\text{PT},\CondQQ{q}\}$. For comparison we include  the most recent lattice determinations. The $J^P\!=\!0^-$ decay constants are taken from  \cite{Aoki:2019cca} which averages over  values in \cite{Bazavov:2017lyh,Gambino:2017vkx,Hughes:2017spc,McNeile:2011ng} and \cite{Yang:2014sea,Bazavov:2011aa,Boyle:2017jwu,Davies:2010ip,Na:2012iu} for the $B$- and $D$-mesons, respectively.  For the $J^P\!=\!1^-$ states we quote the values obtained in \cite{Lubicz:2017asp}.  
   The experimental values are from the PDG review and the extraction of the decay constants 
   involve the CKM matrix $|V_{\textrm{ub}}|$ and $|V_{\textrm{cd(s)}}|$ as inputs. The PDG-error is from the experiment and the CKM input in the first and second parentheses respectively. 
   Note that  the central values for  $f_{B_1}$ and $f_{B_{1s}}$  from  \cite{Wang:2015mxa}     
   deviate considerably from ours which might be due to discrepancies in the $\ORD(\al_s) \vev{\bar qq}$-corrections (cf. remarks at the end of the first paragraph in \SEC \ref{sec:decaySR}).}
  \label{tab:fBparams}
\end{table}

The PDG value, for which the CKM matrix elements $|V_{\textrm{cd(s)}}|$ are inputs, deviates close
to three standard deviations from the lattice result.

\begin{table}[t]
  \centering
  \setlength\extrarowheight{-3pt}
  \begin{tabular}{l| r | r | r | r | r | r | r | r}
                                          & $\fb$              & $\fbS$             & $\fbs$             & $\fbsT$            & $\fd$              & $\fdS$             & $\fds$             & $\fdsT$\\ \hline
   Value                                  & 192.3              & 224.8              & 209.0              & 199.7              & 189.6              & 225.7              & 226.7              & 202.1\\\hline
   Error                                  & $\err{19.7}{18.6}$ & $\err{21.3}{20.3}$ & $\err{22.6}{21.2}$ & $\err{20.7}{19.5}$ & $\err{14.7}{15.4}$ & $\err{17.1}{17.4}$ & $\err{18.3}{16.5}$ & $\err{16.0}{16.3}$\\\hline
   $\Delta s_0^{\fb}$                     & $\err{11.0}{9.5}$  & $\err{12.3}{10.8}$ & $\err{12.0}{10.4}$ & $\err{10.7}{9.2}$  & $\err{10.5}{8.9}$ & $\err{10.8}{9.2}$ & $\err{13.1}{11.1}$ & $\err{11.6}{9.9}$\\
   $\Delta M^2_{\fb}$                     & $\err{0.0}{1.8}$   & $\err{0.0}{2.0}$   & $\err{0.0}{1.0}$   & $\err{0.0}{0.5}$   & $\err{0.0}{1.7}$  & $\err{0.0}{1.6}$  & $\err{0.6}{0.2}$   & $\err{0.8}{0.0}$\\
   $\Delta m_h$                           & $\err{11.4}{11.7}$ & $\err{11.8}{12.0}$ & $\err{14.1}{14.6}$ & $\err{12.9}{13.4}$ & $\err{2.0}{1.9}$  & $\err{2.0}{1.8}$  & $\pm 6.1$          & $\pm 4.4$\\
   $\Delta \vev{\bar{q}q} $               & $\pm 1.7$          & $\pm 1.4$          & $\pm 1.7$          & $\pm 1.8$          & $\pm 2.7$         & $\pm 2.2$         & $\pm 3.0$          & $\pm 2.8$\\
   $\Delta \mukin$                        & $\err{11.2}{10.1}$ & $\err{11.9}{10.8}$ & $\err{12.4}{10.5}$ & $\err{11.2}{9.5}$  & \mc{$-$}          & \mc{$-$}          & \mc{$-$}           & \mcc{$-$}\\
   $\Delta \mu_m$                         & \mc{$-$}           & \mc{$-$}           & \mc{$-$}           & \mc{$-$}           & $\err{4.5}{10.2}$ & $\err{4.4}{11.8}$ & $\err{9.1}{8.1}$   & $\err{6.6}{9.2}$\\
   $\Delta \mu_{\al_s}$                   & $\err{1.8}{3.7}$   & $\err{2.6}{5.2}$   & $\err{1.0}{2.0}$   & $\err{0.9}{1.8}$   & $\err{7.5}{4.4}$  & $\err{11.1}{6.4}$ & $\err{0.7}{1.2}$   & $\err{0.5}{0.9}$\\
   $\Delta \muUV$                         & \mc{$-$}           & \mc{$-$}           & \mc{$-$}           & $\err{1.7}{2.0}$   & \mc{$-$}          & \mc{$-$}          & \mc{$-$}           & $\err{3.2}{3.8}$\\
   $\Delta \frac{\CondQQ{q}}{\CondQQ{s}}$ & \mc{$-$}           & $\pm 2.8$          & \mc{$-$}           & \mc{$-$}           & \mc{$-$}          & $\pm 3.4$         & \mc{$-$}           & \mcc{$-$}\\
   $\Delta_{\textrm{hd}} $                & $\pm 1.0$          & $\pm 0.5$          & $\pm 3.1$          & $\pm 3.7$          & $\pm 3.8$         & $\pm 3.2$         & $\pm 5.4$          & $\pm 5.8$\\
  \end{tabular}
  \caption{\small Breakdown of the main contributions to the uncertainty for a representative selection of decay constants in units of $\MeV$ in the kinetic scheme.  The uncertainty $\Delta_{\textrm{hd}}$ covers   higher dimensional condensates  omitted from the OPE which are estimated as the values of the $d\!=\!4,5$ condensates. 
  This is conservative as the four quark condensates are known to be a sub per mille effect.
The total uncertainty also includes contributions not shown in the table, such as $\Delta m_0^2$ which has a negligible impact.}
  \label{tab:fBErrors}
\end{table}

\subsection{Numerical Analysis}
\label{sec:numf}

The numerical analysis is the same as for the residues/couplings except that 
the scales are taken to be different as, in contrast, there is no motivation to cancel terms 
in ratios. 
Concretely,  the condensate and $\al_s$ scale are changed as shown, to the right of the vertical double
separation, in  \TAB\ref{tab:scales}.
This enforces a change in  SR    parameters  $\{M^2_{f_B}, s_0^{f_B}\}$
according to  the previous criteria, with thresholds fixed such that the daughter SR
\begin{equation}\label{eq:fBdaughterSR}
m_{\Bx}^2=M_{f_{\Bx}}^4\frac{d}{dM_{f_{\Bx}}^2}\ln\int_{m_b^2}^{s^{f_{\Bx}}_0}ds\; e^\frac{-s}{M_{f_{\Bx}}^2} \rho_{f_{\Bx}}(s) \;,
\end{equation}
reproduces the known value of the associated meson mass to  $\approx 0.5\%$. The continuum contribution is kept below $\approx 45\%$. The SR parameters are given alongside the main results in \TAB\ref{tab:fBparams} 
(cf. \TAB\ref{tab:fBparamsMSbar} for $\MSbar$-evaluation of the $B$-meson decay constants) 
and a representative breakdown of the uncertainty is given in \TAB\ref{tab:fBErrors}.  
Isospin breaking effects impact
at the sub per mille level e.g. \cite{Lucha:2016nzv} and are therefore not considered as they are superseded by the actual uncertainties.
If considered, it would seem sensible to include QED effects as well, which would
then necessitate the inclusion of the radiative mode in addition.

The uncertainties of the decay constants are  around $10\%$ and in agreement with 
lattice results   of $\ORD(1$-$4\%)$-uncertainty.  Moreover, we quote other QCD SR determinations,\cite{Gelhausen:2013wia} and  \cite{Lucha:2010ea}. We differ from these results mainly in two aspects. First we do not include partial NNLO effects but  treat the mass scheme and the factorisation scale dependence $\muCond$ separately and thus more 
carefully.
  Secondly, we use a significant  update of the  strange quark condensate. 
 We note that our values are also consistent with 
 the classic Jamin and Lange result \cite{Jamin:2001fw},
 $(\fb ,\fbS)  = ( 210(19),244(21))\MeV$.

\subsubsection{Ratios of Decay Constants}
 \label{sec:ratios}

Some of the decay constants are related by heavy quark and/or  $SU(3)_F$ symmetries,  and thus 
there is some tradition in investigating ratios and determining their deviation from unity. 
A total of  $24$ ratios  are shown in  \TAB\ref{tab:fBRatio} 
(cf. \TAB\ref{tab:fBRatioMSbar} for the $\MSbar$-evaluation of the $B$-meson ratios).

\begin{table}[t]
  \centering
  \setlength\extrarowheight{-4.5pt}
  \begin{tabular}{|  r | r | r | r | r | r | r | r |} \hline\hline
    $\fbS/\fb$   & $\fbsS/\fbs$ & $\fboS/\fbo$  & $\fbsTS/\fbsT$ & $\fboTS/\fboT$ & $\fdS/\fd$      & $\fdsS/\fds$ & $\fdsTS/\fdsT$\\ 
    $1.17(7)$    & $1.17(7)$    & $1.23(8)$     & $1.18(6)$      & $1.24(8)$      & $1.19(7)$       & $1.23(8)$    & $1.27(8)$ \\  \hline  \hline \rule{0pt}{1.3em} 
    $\fbs/\fb$   & $\fbo/\fb$   & $\fbo/\fbs$   & $\fbsT/\fb$    & $\fboT/\fb$    & $\fboT/\fbsT$   & $\fds/\fd$   & $\fdsT/\fd$\\
    $1.09(6)$    & $1.29(9)$    & $1.18(9)$     & $1.04(6)$      & $1.20(9)$      & $1.15(9)$       & $1.20(11)$   & $1.07(9)$ \\  \hline \hline \rule{0pt}{1.3em}
    $\fbsS/\fbS$ & $\fboS/\fbS$ & $\fboS/\fbsS$ & $\fbsTS/\fbS$  & $\fboTS/\fbS$  & $\fboTS/\fbsTS$ & $\fdsS/\fdS$ & $\fdsTS/\fdS$ \\
    $1.09(6)$    & $1.36(8)$    & $1.24(8)$     & $1.05(5)$      & $1.27(8)$      & $1.21(8)$       & $1.23(9)$    & $1.13(7)$ \\ \hline
     \end{tabular}
  \caption{\small 
  Ratios of various decay constants in the kinetic($\MSbar$) scheme for 
  the $B(D)$-mesons. Comparison with the literature 
  can be found in  \SEC\ref{sec:ratios}.
  Ratios in the $\MSbar$ scheme for the $B$-mesons are given in 
   \TAB\ref{tab:fBRatioMSbar}.}
  \label{tab:fBRatio}
\end{table}
 
$SU(3)_F$-type ratios  such as $f_{B_s}/f_B$  are typically above 1
as one would intuitively expect.
\com{We quote our results, denoted by ``PZ" for brevity instead of ``this work", against some results from the literature}
\begin{equation}
\label{eq:SU3}
\left(\frac{\fbS}{\fb},\frac{\fdS}{\fd}\right)    = \left\{  \begin{array}{ll lll}
 ( 1.17(7), & 1.19(7) &\!\!)   & \textrm{PZ}  & \textrm{SRs}\\
 (1.209(5), & 1.174(7)&\!\! ) & \mbox{\cite{Aoki:2019cca}} & \textrm{lattice}  \\ 
 ( -  & 1.265(36) &\!\!)   &  \mbox{\cite{PDG}} & \textrm{experiment}
  \end{array}\right.\;,
\end{equation}
Comparison with the lattice average and  shows 
that there is good agreement albeit the precision in lattice QCD, at the sub per mille level, is beyond reach for 
QCD SRs. The above lattice values are averaged over the works of \cite{Bazavov:2017lyh,Bussone:2016iua,Dowdall:2013tga,Hughes:2017spc} and 
\cite{Na:2012iu,Bazavov:2011aa,Boyle:2017jwu} for the $B$- and $D$-ratio respectively.
The PDG value, for which the CKM matrix elements $|V_{\textrm{cd(s)}}|$ are inputs, deviates close
to three standard deviations from the lattice result.
  Further ratios of interest stem from heavy quark symmetry which groups 
  the $B$ and the $B^*$ meson into the same multiplet  as in this (non-relativistic) 
  limit the spin ceases to matter.
  Deviations of the rations from one therefore highlight sensitivities to effects 
beyond that limit and comparison with the literature 
\begin{equation*}
\def\arraystretch{1.6}
\left(\frac{\fbs}{\fb},\frac{\fbsS}{\fbS},\frac{\fds}{\fd} , \frac{\fdsS}{\fdS} \right)   \! = \!\left\{  \begin{array}{llllll l}
 \!( 1.09(6),              & 1.09(6) ,             & 1.20(11) ,            & 1.23(9)              & \!\!) & \textrm{PZ}                        & \textrm{SRs} \\
 \!( 1.02^{+0.02}_{-0.09}, & 1.04^{+0.01}_{-0.08}, & 1.20^{+0.13}_{-0.07}, & 1.24^{+0.13}_{-0.05} & \!\!) & \mbox{\cite{Gelhausen:2013wia}}    & \textrm{SRs} \\
 \!( 0.944(11)(18) ,       & 0.947(23)(20)  ,      & -                     & -                    & \!\!) & \mbox{\cite{Lucha:2015xua}}        & \textrm{SRs}\\
 \! (  -                    & -                     & -                     & 1.10(2)              & \!\!) & \mbox{\cite{Donald:2013sra}}       & \textrm{lattice} \\
\! (  1.051(17),           & -                     & 1.208(27),            & -                    & \!\!) & \mbox{\cite{becirevic2014insight}} & \textrm{lattice} \\
\!  ( 0.941(26) ,           & 0.953(23)  ,          & -                     & -                    & \!\!) & \mbox{\cite{Colquhoun_2015}}       & \textrm{lattice}   \\
\! (  0.958(22),           & 0.974(10),            & 1.078(36),            & 1.087(20)            & \!\!) & \mbox{\cite{Lubicz:2017asp}}       & \textrm{lattice}
\end{array}\right.
\end{equation*}
does show some minor tension between the results.  Note that the lattice result 
\cite{becirevic2014insight} is with $N_f =2 $ and   \cite{Colquhoun_2015,Lubicz:2017asp} 
are with $N_f = 2+1$ and thus more reliable. 
For further discussion of the possible reasons for discrepancies cf.  section IV in \cite{Colquhoun_2015}.
 
 We now proceed to give some detail on of the individual uncertainties of the ratios 
 in the SR computation. 
 In both the $B$- and $D$-meson rations the effective thresholds prove to be the largest source of uncertainty. Whilst correlations between the thresholds, discussed previously, act to constrain the error the contribution to the total uncertainty is still significant, sitting in the region of $\approx70 $-$80\%$. The remaining uncertainty can be mostly attributed to the associated quark mass and in the $D$-meson rations the coupling scale $\mu_{\al_s}$ provides a contribution to the total uncertainty of a similar order. 
 For the $SU(3)_F$-ratios in \eqref{eq:SU3}, the quark condensate ratio  $\CondQQ{s}/\CondQQ{q}$ provides a notable contribution to the total uncertainty.

\section{Summary and Discussion}
\label{sec:conc}

In this work we have determined the 
couplings of photons to heavy-light quark mesons \eqref{eq:Leff} 
 from light-cone sum rules at next-to-leading order in $\al_s$ 
at the twist-$1$,-$2$ level, at leading order in twist-$3$, and partial twist-$4$.\footnote{We have argued 
(cf. sec 3.3. in \cite{Janowski:2021yvz}) that most twist-$4$ parameters require the inclusion of $4$-particle 
distribution amplitudes which have not been classified to date. This can be seen from the equation 
of motion for the form factors not closing or by writing down the $4$-particle distribution amplitude 
of twist-$4$ and subjecting it to the equation of motion of distribution amplitudes.} We have also investigated the effect of various duality regions (cf. \SEC\ref{sec:duality} and \TAB\ref{tab:gValuesA}) and have found the impact to be small.
Our main results, with uncertainties of $\ORD(15\%)$, are given in \TAB\ref{tab:gValues} 
along other theoretical and experimental results for comparison.
  The residues related to the $\bar{B} \to \ga$ form factors, as in 
\eqref{eq:VFF} and \eqref{eq:r}, are given in \TAB\ref{tab:rValues}.
As a by-product  we have determined 
the heavy decay constants  
$f_H,f_{H^*},f_{H_1},f_{H^*}^T$ and $f_{H_1}^T$ ($H = B,D$) in QCD sum rules at next-to-leading 
order.\footnote{With the exception of the $D_1$ as it is not well isolated cf. footnote \ref{foot:D1}.}  
To the best of our knowledge  $\{f_{B^*(D^*)}^T, f_{B_1}^T\}$  have not been evaluated 
with QCD sum rules and we therefore close a gap in the literature.  
Agreement is found with existing results, where comparison is possible, 
on the  analytic and  numerical 
level cf. \TAB\ref{tab:fBparams}.  Our treatment differs, besides
a significant update to the strange quark condensate,  in that 
we treat the mass-scheme and the factorisation scale dependence $\muCond$ separately and thus more 
carefully, but do not include partial $\ORD(\al_s^2)$ corrections to perturbation theory.  Ratios of decay constant 
are given in \TAB\ref{tab:fBRatio} and compared to the literature in \SEC\ref{sec:ratios}. 
\begin{table}[H]
  \centering 
  \setlength\extrarowheight{-4.5pt}
 \begin{tabular}{| l |  l | l | } \hline\hline
$\Gamma(B^{*0}\!\to\! B^0\gamma)$   & $\Gamma(B_s^{*}\!\to\! B_s\gamma)$ & $\Gamma(B^{*+}\!\to\! B^+\gamma)$   \\ 
$0.16^{+0.06}_{-0.06}\keV$          & $0.24^{+0.08}_{-0.08}\keV$         & $0.45^{+0.17}_{-0.16}\keV$  \\ \hline \hline
\rule{0pt}{1.3em}
$ \Gamma(B^{0}_1\!\to\! B^0\gamma)$ & $\Gamma(B_{1s}\!\to\! B_s\gamma)$  & $\Gamma(B^{+}_1\!\to\! B^+\gamma)$   \\
$26.22^{+17.00}_{-17.00}\keV$       & $ 41.14^{+22.44}_{-22.44}\keV$       & $99.30^{+57.92}_{-55.16}\keV$  \\ \hline
\hline 
\rule{0pt}{1.3em} 
$\Gamma(D^{*0}\!\to\! D^0\gamma)$   & $\Gamma(D_s^{*}\!\to\! D_s\gamma)$ & $\Gamma(D^{*+}\!\to\! D^+\gamma)$   \\
$27.83^{+9.23}_{-9.50}\keV$       & $ 2.36^{+1.49}_{-1.41}\keV$        & $0.96^{+0.58}_{-0.62}\keV$  \\ \hline
\end{tabular}
\caption{\small Decay rates based on the $g$-couplings in \TAB\ref{tab:gValues} and the decay rate formula \eqref{eq:widths}.}
\label{tab:rates}
\end{table}
We now turn to phenomenological aspects. 
The coupling determinations lead to 
the radiative decay  predictions given in \TAB\ref{tab:rates}, consistent with the experimentally known
 $D^{+}/D^0$-rates. 
 It's unfortunate that the $B$-rates are not experimentally known as
 our predictions are more reliable in that sector  
 (e.g.  independence of the interpolating current and convergence of the twist expansion). 
 Particularly for the $D^+/D_s$-channels there is the additional issue of large cancelation of the
$Q_c$- and  $Q_{q/s}$-contributions which present a challenge for all theory approaches (cf. the discussion in 
\SEC\ref{sec:gcomp}).  An important aspect is the interplay with the real QED-corrections 
in leptonic decays $H \to \ell \bar{\nu}(\ga)$.
This is the case  since the couplings describe the pole residue \eqref{eq:VFF} and  \cite{Becirevic:2009aq} which, bearing in mind previously  mentioned cancellations, should play a significant role in the soft-photon emission. 
In view of the importance of QED-corrections at the precision frontier, these couplings will hopefully 
attract further attention from the experimental and theory community.

\subsection*{Acknowledgments}

RZ is supported by an STFC Consolidated Grant, ST/P0000630/1. BP is supported by an STFC Training Grant, ST/N504051/1. 
We are grateful to Marco Pappagallo, Christine Davies, Giuseppe Gagliardi, Christopher Sachrajda  for useful discussions and to James Gratrex for thorough proofreading of the manuscript.

\appendix

\section{Convention, Definitions and Additional Tables}
\label{app:conv}

In this appendix we collect conventions, definitions and input parameters. 
\subsection{Convention and Definitions}

We use the convention $\varepsilon_{0123} = 1$ for the Levi-Civita tensor and
$D_\mu = \partial_\mu + s_e i e Q_f A_\mu + s_g i g_ s A_\mu$ for the 
covariant derivative ($e > 0$ and $Q_e = -1$ for the electron as a $u$-spinor). 
Below we will keep explicit factors  $s_i$ in place, which are assumed  $s_i = 1$  throughout 
the main text, in order to facilitate comparison with the literature.
The $B_q$-meson $(q = d,u,s)$ decay constant is defined by 
\begin{equation}
\label{eq:fB}
\matel {0 }{ \bar{q}  \gamma^{\mu} \gamma_{5}  b  }{ \bar{B}_q (p_B) } 
= s_B i p_B^{\mu} f_{B_q} \;,
\end{equation}
 and for the $B^*_q$ $(1^-)$ and $\Bqone$ $(1^+)$ states via
\begin{alignat}{4}
\label{eq:fBs}
& \matel{0} { \bar{q} \, \gamma^{\mu}  \, b}{ {\bar{B}_q^*}(p)} &\;=\;& s_{B^*} m_{B_q^*}  f_{B_q^*} \eta^{\mu}  \;, \qquad\;  & & \matel{0} { \bar{q} \, \gamma^{\mu}  \ga_5 \, b}{ \bar{B}_{1q} (p)} &\;=\;& m_{\Bqone} s_{B_1} f_{\Bqone} \eta^{\mu}  \;, \nonumber \\[0.1cm] 
& \matel{0} { \bar{q} \, \sigma^{\mu\nu}  \, b}{ {\bar{B}_q^*}(p)} &\;=\;& i s_{B^*} f_{B_q^*}^T   \eta^{[\mu} p^{\nu]}  \;,\qquad  & & \matel{0} { \bar{q} \, \sigma^{\mu\nu}  \ga_5 \, b}{ {\bar{B}_{1q}}(p)} &\;=\;& -i  s_{B_1} f_{\Bqone}^T   \eta^{[\mu} p^{\nu]}  \;.
\end{alignat}
The definition for the $D-$, $D^*$- and $D_1$-mesons are analogous. 
With these conventions the couplings the effective Lagrangian \eqref{eq:Leff}  assumes the form
\begin{equation}
\label{eq:Leffs}
{\cal L}_{\textrm eff} =  s_e s_{B} s_{B^*}  \frac{1}{2} \gonemi \, \epsilon(B^*,\partial B^\dagger,F) - i 
s_e s_{B} s_{B_1}  \gonepl \, B_{1\al} \partial_\be B^\dagger F^{\al \be}+ \textrm{h.c.}   \;.
\end{equation}
For completeness we state the definition of the $B \to \ga$ form factors used in \cite{Janowski:2021yvz}
\begin{alignat}{3}
\label{eq:ffs}
& \matel{ \gamma(k,\epsilon) }{O^V_\mu }{ \bar{B}_q (p_B) } 
 &\;=\;&   s_B s_e (  \Pperp{\mu} \, \FV (q^2)  &\;-\;&  \Ppara{\mu} \, \left(  \FA(q^2)  + 
Q_{\bar{B}_q}  \frac{2 f_{B_q}/m_{B_q} }{1-q^2/m_{B_q}^2} + \dots  \right) \;, \nonumber \\[0.1cm]
& \matel{ \gamma(k,\eps) }{ O^T_\mu  }{ \bar{B}_q (p_B) }
 &\;=\;&  s_B s_e(  \Pperp{\mu }   \, \FTV(q^2) &\; -\;& \Ppara{\mu }  \, \FTA(q^2) ) \;,
 \end{alignat}
 where $\Pperp{\mu}$ and $\Ppara{\mu}$ are defined 
 in the main text \eqref{eq:P}, $Q_{\bar{B}_q} $ is the $\bar B$-meson charge  and 
 the dots represents  the Low-term (or contact term) 
 which is not important for this paper (cf. \cite{Janowski:2021yvz} for details). 
 Note that, the point-like term, proportional to $f_B$, is not be included for the $g_{ B_{u} B_{1u}  \ga}$ coupling as it is not associated with the $B_{1u}$-pole. 
 The local operators in \eqref{eq:ffs}  are given by $O^{V[T]}_{\mu}\equiv O^{V[T]}_{\perp\mu}+O^{V[T]}_{\parallel\mu}$, with
\begin{align}\label{eq:TVop}
  O^V_{\perp\mu} &\equiv -  \frac{1}{e} m_{B_q} \bar{q} \gamma_{\mu} b \;, \qquad O^T_{\perp\mu}  \equiv \frac{1}{e}  \bar{q} i q^\nu \sigma_{\mu \nu} b \;,\nonumber\\[0.1cm]
  O^V_{\parallel\mu} &\equiv   \frac{1}{e} m_{B_q} \bar{q} \gamma_{\mu}  \ga_5 b \;, \qquad O^T_{\parallel\mu}  \equiv \frac{1}{e}  \bar{q} i q^\nu \sigma_{\mu \nu} \ga_5 b \;.
\end{align}

\subsection{Additional Tables}
\label{app:tables}
\FloatBarrier
Here we provide some additional tables, namely the input parameters \TAB\ref{tab:inputParams}, $\MSbar$ determinations of the decay constants \TAB\ref{tab:fBparamsMSbar} and their ratios \TAB\ref{tab:fBRatioMSbar}. 
\begin{table}[btp]
  \centering
  \setlength\extrarowheight{-4.5pt}
  \resizebox{\columnwidth}{!}{
  \begin{tabular}{| l | ll  | ll | ll |} \hline\hline
     $\!\!H(\MeV)$                          & $B(5280)$            & $B_s(5367)$          & $B^*(5325)$          & $B^*_s(5415)$        & $B_1(5726)$          & $B_{1s}(5829)$\\ \hline
      \rowcolor{gray!10}
     $\!\!f_{\Bx}$                          & $\Verr{213}{22}{16}$ & $\Verr{248}{23}{17}$ & $\Verr{218}{23}{27}$ & $\Verr{260}{23}{22}$ & $\Verr{288}{25}{24}$ & $\Verr{341}{20}{24}$\\
     $\delta_\text{PT},\;\delta_\CondQQ{q}$ & -0.03, -0.11         & 0.03, -0.06          & -0.18, -0.23         & -0.12, -0.22         & -0.11, -0.19         & -0.05, -0.13\\
     $s_0^{f_{\Bx}}$, $M^2_{f_{\Bx}}$       & 33.6, 6.0            & 34.9, 7.2            & 33.7, 6.5            & 35.0, 6.8            & 39.0, 7.5            & 40.8, 9.4\\ \hline \hline  \rule{0pt}{1.3em}
     $\!\!H(\MeV)$                          &                      &                      & $B^*(5325)^T$        & $B^*_s(5415)^T$      & $B_1(5726)^T$        & $B_{1s}(5829)^T$  \\ \hline
      \rowcolor{gray!10}
     $\!\!f^T_{\Bx}$                        &                      &                      & $\Verr{208}{21}{23}$ & $\Verr{249}{20}{19}$ & $\Verr{267}{21}{22}$ & $\Verr{318}{18}{22}$ \\
     $\delta_\text{PT},\;\delta_\CondQQ{q}$ &                      &                      & -0.16, -0.24         & -0.11, -0.24         & -0.09, -0.19         & -0.04, -0.14\\
     $s_0^{f_{\Bx}}$, $M^2_{f_{\Bx}}$       &                      &                      & 33.7, 6.5            & 35.0, 6.8            & 39.0, 6.2            & 40.8, 9.4 \\ \hline
  \end{tabular}
  }
  \caption{\small  B-meson decay constants, $\MeV$-units, determined in the $\MSbar$ scheme (kinetic scheme values in \TAB\ref{tab:fBparams}))
 with the Borel parameter and effective threshold given in $\GeV^2$-units.}
  \label{tab:fBparamsMSbar}
\end{table}

We note that when fixing the SR parameters via the daughter SR we observe that the optimal value of the effective threshold for the $B^*$ decay constant sits below that of the $B$. Clearly this does not make sense from a physical point of view and so we relax the condition on the daughter SR \eqref{eq:fBdaughterSR} such that it reproduces the associated meson mass to within $1.5\%$, which allows for the physical ordering of the thresholds to be imposed. We do not observe this problem when evaluating in the kinetic scheme which is another reason in its favour.
\begin{table}[btp]
  \centering
  \setlength\extrarowheight{-4.5pt}
  \begin{tabular}{|  r | r | r | r | r | r |} \hline\hline
    $\fbS/\fb$               & $\fbsS/\fbs$             & $\fboS/\fbo$             & $\fbsTS/\fbsT$           & $\fboTS/\fboT$           & \\  
    $1.17^{+0.07}_{-0.07}$ & $1.18^{+0.08}_{-0.08}$ & $1.18^{+0.08}_{-0.06}$ & $1.19^{+0.07}_{-0.07}$ & $1.18^{+0.08}_{-0.05}$ & \\ \hline\hline  \rule{0pt}{1.3em}
    $\fbs/\fb$               & $\fbo/\fb$               & $\fbo/\fbs$              & $\fbsT/\fb$              & $\fboT/\fb$              & $\fboT/\fbsT$   \\ 
    $1.04^{+0.13}_{-0.08}$ & $1.35^{+0.12}_{-0.08}$ & $1.30^{+0.09}_{-0.10}$ & $0.98^{+0.11}_{-0.07}$ & $1.26^{+0.09}_{-0.07}$ & $1.28^{+0.08}_{-0.10}$        \\  \hline\hline \rule{0pt}{1.3em}
    $\fbsS/\fbS$             & $\fboS/\fbS$             & $\fboS/\fbsS$            & $\fbsTS/\fbS$            & $\fboTS/\fbS$            & $\fboTS/\fbsTS$  \\ 
    $1.05^{+0.12}_{-0.07}$ & $1.37^{+0.09}_{-0.08}$  & $1.31^{+0.08}_{-0.11}$ & $1.00^{+0.11}_{-0.07}$  & $1.28^{+0.08}_{-0.07}$ & $1.27^{+0.10}_{-0.08}$  \\  \hline
     \end{tabular}
  \caption{\small 
  Ratios of  decay constants in the $\MSbar$ scheme. The asymmetry of the uncertainty,  most pronounced in the pseudo-scalar vs. vector vs. tensor channels, arises from an  asymmetric variation of the $\MSbar$ scale, cf. \TAB\ref{tab:scales}. 
  The corresponding ratios in the kinetic scheme are given in  \TAB\ref{tab:fBRatio}
  which are compatible within uncertainties.}
  \label{tab:fBRatioMSbar}
\end{table}

\begin{table}[btp]
\addtolength{\arraycolsep}{3pt}
\renewcommand{\MeV}{\,{\textrm{MeV}}}
\renewcommand{\GeV}{\,{\textrm{GeV}}}
\renewcommand{\arraystretch}{1.3}
\resizebox{\columnwidth}{!}{
\begin{tabular}{C|C|C|C|C|C} 
\multicolumn{6}{c}{\mbox{Running coupling parameters}}\\\hline
\alpha_s(m_Z) \mbox{~\cite{PDG}} & m_Z\mbox{~\cite{PDG}}\\\hline
0.1176(20) & 91.19 \GeV \\\hline 
\multicolumn{6}{C}{J^P =0^- \mbox{ Meson masses~\cite{PDG}}}\\\hline
\mBz        & \mBu       & \mBs       & \mDz       & \mDd       & \mDs   \\\hline
5.280  \GeV & 5.280 \GeV & 5.367 \GeV & 1.865 \GeV & 1.870 \GeV & 1.968 \GeV \\\hline
\multicolumn{6}{C}{J^P =1^- \mbox{ Meson masses~\cite{PDG}}}\\\hline
\mBsz       & \mBsu      & \mBss      & \mDsz      & \mDsd      & \mDss   \\\hline
5.325  \GeV & 5.325 \GeV & 5.415 \GeV & 2.007 \GeV & 2.010 \GeV & 2.112 \GeV \\\hline
\multicolumn{6}{C}{J^P =1^+ \mbox{ Meson masses~\cite{PDG}}}\\\hline
\mBoz       & \mBou      & \mBos      & \mDoz      & \mDod      & \mDos   \\\hline
5.726  \GeV & 5.726 \GeV & 5.829 \GeV & 2.421 \GeV & 2.423 \GeV & 2.460 \GeV \\\hline

 \multicolumn{6}{C}{\mbox{Quark masses~\cite{PDG}}}\\\hline
m_{s\lscale{2}}            & m_b(m_b)     & m_c(m_c)       & m_b^{\textrm{pole}} & m_c^{\textrm{pole}} & m_b^{kin}(1\GeV)^{\dagger}  \\\hline
92.9(7)  \MeV                & 4.18(4) \GeV & 1.27(2) \GeV   & 4.78(6)\GeV        & 1.67(7) \GeV        & 4.53(6)\GeV  \\\hline

 \multicolumn{6}{C}{\mbox{Condensates}}\\\hline

\qbarq_\lscale{2} \mbox{~\cite{Bali:2012jv}} & \sbars \mbox{~\cite{McNeile:2012xh}} & \vev{G^2}\mbox{~\cite{SVZ79I,SVZ79II}} & m_0^2 \mbox{~\cite{Ioffe:2002ee}} & & \\\hline
-(269(2) \MeV)^3                             & 1.08(16) \qbarq                      & 0.012(4)\GeV^4                         & 0.8(2)\GeV^2                      & & \\\hline
\end{tabular}
}
\caption{\small Summary of input parameters. 
$^{\dagger}$ Value obtained by using the $\ORD(\al_s^2)$ conversion between the 
$\MSbar$ and the kinetic mass  given in \cite{Gambino:2017vkx}. 
The uncertainty is obtained by  adding in quadrature the uncertainty due to the $\MSbar$ mass and the conversion formula.  
For the meson masses we have not indicated an uncertainty as they are negligible.
We refer to \cite{Janowski:2021yvz} for all the input concerning the photon DA that enters the light-cone sum rule computation.
}
\label{tab:inputParams}
\end{table}
\FloatBarrier
\section{Analytic Results for the \texorpdfstring{$f_H,f_{H^*},f_{H_1},f_{H^*}^T$}{} and \texorpdfstring{$f_{H_1}^T$}{} Decay Constants}
\label{app:decayConsts}

In this appendix we provide the analytic results for the decay constants $\{f_B,f_{B^*},f_{B_1},f_{B^*}^T,f_{B_1}^T\}$, with straightforward substitutions for the  
 their $D$-meson counterparts.  \com{The $f_{B^*}^T,f_{B_1}^T$ results are new
 and comparison with the literature with regards to $f_B,f_{B^*},f_{B_1}$  is commented on at the 
 beginning of \SEC\ref{sec:decaySR}.}
 We give the results in terms of the densities $\rho_{f_{\Bx}}(s)$  
 and Wilson coefficients $c_j^{f_{\Bx}}$ that enter \eqref{eq:fBx}. 
 The densities  are related to the correlation functions as follows 
 \begin{equation}
  \rho_{f_B}(s) = \frac{1}{\pi} \frac{\textrm{Im}_s \Ga_{\fb}(s)}{(m_b+m_q)^2 } \;, \quad 
   \rho_{f_{B^*}}(s) = \frac{1}{\pi} \textrm{Im}_s \Ga_{\fbs}(s) \;,\quad 
    \rho_{f^T_{B^*}}(s) = \frac{1}{\pi} \textrm{Im}_s \Ga_{\fbsT}(s) \;.
  \end{equation}
 The Wilson coefficients are presented after integration and can therefore depend on the effective threshold.   For comparison with the literature cf. footnote \ref{foot:compare} in the main text.
 
The leading contribution to the local OPE is the perturbative one which we further decompose into LO and NLO parts
\begin{equation}
\rho(s)=\;\rho^{(0)}(s)+\frac{\al_s}{\pi}\rho^{(1)}(s)+\dots \;. 
\end{equation}
At LO, including corrections due to the light quark mass to $\ORD(m_q^2)$, we find
\begin{alignat}{1}
\rho^{(0)}_{\fb}(s)=\;   & \frac{N_c}{8\pi^2}\,s\left(\bar z^2+2\,\frac{m_q}{m_b}z\bar z-2\frac{m_q^2}{m_b^2}z\right),\nonumber\\[0.1cm]
\rho^{(0)}_{\fbs}(s)=\;  & \frac{N_c}{24\pi^2}s\left(\bar z^2(z+2)+6\,\frac{m_q}{m_b}z\bar z-3\frac{m_q^2}{m_b^2}z(z^2+1)\right),\nonumber\\[0.1cm]
\rho^{(0)}_{\fbsT}(s)=\; & \frac{N_c}{24\pi^2}s\left(\bar z^2(2z+1)+6\frac{m_q}{m_b}z\bar z-6\frac{m_q^2}{m_b^2}z^3\right),
\end{alignat}
whilst at NLO,
\begin{alignat}{1}
\rho^{(1)}_{\fb}(s)=\;   & \frac{N_c C_F}{16\pi^2}s\bar z\Bigg[\frac{9}{2}\bar z+(z-3)(2z-1)\ln z +\bar z(2z-5+2\ln z)\ln\bar z + 4\bar z\text{Li}_2(z) -(3z-1)r_S\Bigg],\nonumber\\[0.3cm]
\rho^{(1)}_{\fbs}(s)=\;  & \frac{N_c C_F}{16\pi^2}s\Bigg[1-\frac{5}{2}z+\frac{2}{3}z^2+\frac{5}{6}z^3+\frac{1}{3}z(5z^2-4z-5)\ln z -\frac{1}{3}\bar z^2(5z+4-2(z+2)\ln z)\ln\bar z\nonumber\\[0.1cm]
                         & +\frac{4}{3}\bar z^2(z+2)\text{Li}_2(z)+z(z^2-1) r_S \Bigg],\nonumber \\[0.3cm]
\rho^{(1)}_{\fbsT}(s)=\; & \frac{N_c C_F}{96\pi^2}s\Bigg[\frac{7}{3}+2z-15z^2+\frac{32}{3}z^3+2(8z^3-11z^2+2z-1)\ln z 
+8\bar z^2(2z+1)\text{Li}_2(z)
 \nonumber\\[0.1cm]
&   -2\bar z^2(8z+1-2(2z+1)\ln z)\ln\bar z-2\bar{z}^2(2z+1)\logMu{UV} -12z^2\bar zr_S\Bigg] \;,
\end{alignat}
with the $\ORD(m_q)$ corrections given by, 
\begin{alignat}{1}
\delta_{m_q}\rho^{(1)}_{\fb}(s)=\; & \frac{N_c C_F}{4\pi^2}\frac{m_q}{m_b}s\Bigg[3z-3z^2+z(z^2-5z+3)\ln z+z\bar z(z-{4}+\ln z)\ln\bar z\nonumber \\[0.1cm]
                                     & +2z\bar z\text{Li}_2(z)-\frac{1}{2}z(3z-2) r_S \Bigg],\nonumber \\[0.3cm]
\delta_{m_q}\rho^{(1)}_{\fbs}(s)=\; & \frac{N_c C_F}{8\pi^2}\frac{m_q}{m_b}s\Bigg[\frac{9}{2}z-5z^2+\frac{1}{2}z^3-z(z^2+4z-3)\ln z-z\bar z(z+5-2\ln z)\ln\bar z\nonumber \\[0.1cm]
                                     & +4z\bar z\text{Li}_2(z)-z(2z-1)r_S\Bigg],\nonumber \\[0.3cm]
\delta_{m_q}\rho^{(1)}_{\fbsT}(s)=\; & \frac{N_c C_F}{4\pi^2}\frac{m_q}{m_b}s\Bigg[3z-3z^2-z(z^2+z-1)\ln z-z\bar z(z+2-\ln z)\ln\bar z\nonumber \\[0.1cm]
                                     & +2z\bar z\text{Li}_2(z)+\frac{1}{4}\bar z z\logMu{UV}+\frac{1}{4}z(1-3z)r_S\Bigg] \;,
\end{alignat}
where $z \equiv m_b^2/s$.  The $\muUV$ dependence is  consistent with 
the anomalous scaling \eqref{eq:gaT}.

The Borel subtracted non-perturbative contributions are given by,
\begin{alignat}{2}
\label{eq:B6}
c_{\CondQQ{q}}^{\fb}=    & -\Bigg[1-\frac{m_q}{2m_b}-\frac{m_b\,m_q}{2M^2}\nonumber\\
&+\frac{\al_s C_F}{2\pi}\bigg\{1 +3 \Ga_{0} - \frac{3}{2}\logMu{cond}+ \left(\frac{3}{2}-\frac{m_b^2}{M^2} \right)r_S \bigg\} \Bigg]\;,\nonumber\\[0.3cm]
c_{\CondQQ{q}}^{\fbs}=   & -\Bigg[1-\frac{m_b\,m_q}{2M^2}\nonumber\\
&+\frac{\al_s C_F}{2\pi}\bigg\{-1-\frac{m_b^2}{M^2} \Ga_{-1}- \frac{3}{2}\logMu{cond}
+\left(\frac{1}{2}-\frac{m_b^2}{M^2}\right)r_S   \bigg\}\Bigg],\nonumber\\[0.3cm]
c_{\CondQQ{q}}^{\fbsT}=  & -\Bigg[1+\frac{m_q}{2m_b}-\frac{m_b m_q}{2M^2}\nonumber\\
&+\frac{\al_s C_F}{2\pi}\bigg\{-1 - \Ga_0-\logMu{UV}-\frac{3}{2}\logMu{cond}+\left(\frac{1}{2}-\frac{m_b^2}{M^2}\right)r_S\bigg\} \Bigg],\nonumber\\[0.3cm]
\end{alignat}
\begin{alignat}{4}
& c_{\CondG}^{\fb}      & = & \frac{1}{12} \;,                                   & \quad c_{\CondG}^{\fbs}&=      -   \frac{1}{12} \;,     & \quad c_{\CondG}^{\fbsT}&=       -\frac{1}{12}\left(1 +\frac{2m_b^2}{M^2} \Ga_{-1}\right) \;, \nonumber\\[0.2cm]
& c_{\CondqGq{q}}^{\fb} & = & -\frac{1}{2 }\left(1-\frac{m_b^2}{2M^2}\right) \;, & \quad c_{\CondqGq{q}}^{\fbs}&=   \frac{m_b^2}{4 M^2}\;, & \quad c_{\CondqGq{q}}^{\fbsT}&=  \frac{1}{6 }\left(1+\frac{3m_b^2}{2M^2}\right)\;,
\end{alignat}
where the Borel parameter $M^2 \to M^2_{f_{\Bx}}$ accordingly, and
\begin{equation}
\Ga_k = e^{\frac{m_b^2}{M^2}} \left(\Gamma\left(k,\frac{s_0}{M^2}\right)-\Gamma\left(k,\frac{m_b^2}{M^2}\right)\right)\;,
\end{equation}
with $\Gamma(n,z)=\int_z^{\infty} dt\,t^{n-1}e^{-t}$ denoting the  incomplete gamma function. The quantity
\begin{equation}
r_S =   \left\{  \begin{array}{ll}  
 3\ln\left(\frac{\mu_{m}^2}{m_b^2}\right)+4 \qquad \qquad & \MSbar \\
 0 & \textrm{pole} \\
  \frac{16}{3}  \frac{\mu_{\textrm{kin}}}{m_b} + 2\frac{ \mu_{\textrm{kin}}^2}{m_b^2}  & \textrm{kinetic}
\end{array} \right. \;,
\end{equation}
 is a factor that depends on the mass scheme. Above we have also included the leading light quark mass corrections to the LO quark condensate contribution. As mentioned in \SEC\ref{sec:fBcomp},  
 we have verified that the NLO scale dependence, in $\muUV$ and $\mu_{\textrm{cond}}$, is consistent with the LO expression.
   
The SRs for the $J^P=1^+$ decay constants can be obtained from the $J^P=1^-$ ones by changing the sign of certain contributions according to their chirality,
\begin{alignat}{4}
\label{eq:doubling}
& \rho_{f_{B_1}^{(T)}}           & \; =\;                                 & \rho_{f_{B^*}^{(T)}} \;, \qquad
&                                   & c_{\CondQQ{q}}^{f_{B_1}^{(T)}}  & \; =\;                                      & - c_{\CondQQ{q}}^{f_{B^*}^{(T)}}\;,  \nonumber \\[0.1cm]
& c_{\CondG}^{f_{B_1}^{(T)}} & \;=\;                                  & c_{\CondG}^{f_{B^*}^{(T)}}\;,  \quad
&                                   & c_{\CondqGq{q}}^{f_{B_1}^{(T)}} & \; =\;                                      & -  c_{\CondqGq{q}}^{f_{B^*}^{(T)}} \;,
\end{alignat}
in spirit with the parity doubling proposal in \cite{Gratrex:2018gmm}.

\section{Double Dispersion Relation}
\label{app:double}

In computing the densities we are faced with the following problem. We have an analytic function $F(p_B^2,q^2)$ 
for which it is straightforward to derive a single dispersion relation
\begin{equation}
F(p_B^2,q^2) = \int_{m_b^2}^\infty ds \frac{\rho(s,q^2)}{s-p_B^2-i0} \;,
\end{equation}
where  the density is formally given by  $\pi \rho(s,q^2) = \Ima_s F(s,q^2)$. The density  
can be decomposed into poles in $s = q^2$ such that
 \begin{equation}\label{eq:single_disp}
 F(p_B^2,q^2) =\sum_{n\geq0}F_n(p_B^2,q^2)\;,\qquad F_n(p_B^2,q^2)=\int_{m_b^2}^{\infty} ds\frac{\rho_n(s,q^2)}{(s-p_B^2-i0)(s-q^2)^n}\;.
 \end{equation}
 The singularities in $s=q^2$ are of so-called second type, 
which are special solutions of the Landau equations
 \cite{Itzykson:1980rh, Zwicky:2016lka}.
It is our task to write  the $q^2$-dependence of \eqref{eq:single_disp} dispersively, say in an integral over $dt$, and impose a continuum subtraction.  The duality interval is discussed in \eqref{eq:t0} in the main text.
\subsection{Leading Order}
At LO in PT the $\rho_i$ themselves contain no non-trivial cuts. Consequently, the poles provide the only contribution to the discontinuity in $q^2$, allowing us to write 
\begin{equation}
  F_n(p_B^2,q^2)= \frac{1}{\Gamma(n)}\int_{m_b^2}^{\bsup{m_b^2}} ds\int_{m_b^2}^{\btup{s}}\!\! dt\,\frac{\rho_n(s,t)\delta^{(n-1)}(t-s)}{(s-p_B^2)(t-q^2)}\;,
\end{equation}
where the continuum subtraction has been implemented as in \eqref{eq:t0}. Partially integrating and performing the integrals over the $\delta$-functions we obtain,
 \begin{align}\label{eq:PTLO_double_borel}
  F_n(p_B^2,q^2)=\;&\frac{(-1)^{n-1}}{\Gamma(n)}\int_{m_b^2}^{\Sred} ds\,\partial_t^{(n-1)}\big(\rho_n(s,t)g(s,t)\big)\big|_{t\to s}\nonumber\\
  +\;&\frac{1}{\Gamma(n)}\sum_{\ell=1}^{n-1}(-1)^{\ell-1}\left(\frac{\tilde{s}_0^a}{\tilde{s}_0^a+\tilde{t}_0^a}\right)^{n-\ell}\partial_s^{(n-\ell-1)}\left(\partial_t^{(\ell-1)}[\rho_n(s,t)g(s,t)]\big|_{t\to\btup{s}}\right)\bigg|_{s\to\Sred}\;,
 \end{align}
with $g(s,t)=1/(s-p_B^2-i0)(t-q^2-i0)$. The double Borel transform can then be trivially computed by taking $g(s,t)\to\hat{g}(s,t)=e^{-s/M_1^2-t/M_2^2}$.
\subsection{Next-to-Leading Order}
At NLO the situation is complicated by $\rho_n(s,q^2)$ containing polylogarithmic terms that contribute to the discontinuity in $q^2$ in addition to the poles. To lessen this complication only provide a derivation of the double dispersion relation for $n\leq3$,
which is sufficient for the case at hand where the density can be decomposed as
\begin{equation}
\rho(s,q^2) = \rho_0(s,q^2) + \frac{\rho_1(s,q^2)}{(s-q^2)}  + \frac{\rho_2(s,q^2)}{(s-q^2)^2} 
+ \frac{\rho_3(s,q^2)}{(s-q^2)^3} \;.
\end{equation}
Without committing to a specific value for the parameter $a$, we obtain formally a double dispersion relation, with continuum subtraction as in \eqref{eq:t0},  
\begin{align}\label{eq:double_disp_2}
F^{\tzerot}_{\szerot} (p_B^2,q^2) = & \int_{m_b^2}^{\btup{m_b^2}} \!\!\frac{dt}{t-q^2} 
P_s \!\int_{m_b^2}^{\bsup{t}} \!\!  \frac{ds}{s-p_B^2} \hat\rho(s,t)   +   
 P_{\Sred}\!\int_{m_b^2}^{\btup{m_b^2}} \!\!\!\frac{dt}{t-q^2} \, \bar{\rho}(t,p_B^2,q^2,\szerot,\tzerot)\nonumber\\[0.2cm]
& +\int_{m_b^2}^{\Sred}\frac{ds}{s-p_B^2}\tilde{\rho}(s,q^2)+ C( p_B^2,q^2,\szerot,\tzerot)\;,
\end{align} 
where $F^{\tzerot}_{\szerot} \to F  $ as $\szerot,\tzerot\to \infty$. The function $\bar{\rho}(t,p_B^2,q^2,\szerot,\tzerot)$ arises from  partial integration in $s$ in order to reduce the integrands to simple  $1/(s-t)$-poles.
The natural order of integration has been reversed in an attempt to remove complications at the lower integration boundary when integrating-by-parts. The order-1 poles, hidden in $ \hat \rho(s,t)$ and $ \bar\rho(t,p_B^2,q^2,\szerot,\tzerot)$, are handled with the principle part prescription, with $P_x$ denoting the principal value w.r.t. to $1/(x-t)$.
In terms of $\rho_i$, the above functions read
\begin{alignat}{2}
& \hat\rho(s,t)         & \;=\; & \frac{1}{\pi}\left( \Ima_t   \rho_0  \pl \frac{1}{s\mi t} \left[  \Ima_t {\rho}_1
\mi
(s-p_B^2) \left( \left( \frac{\Ima_t {\rho}_2}{s-p_B^2}  \right)^{'}  + \frac{1}{2} \left( \frac{\Ima_t {\rho}_3}{s-p_B^2}\right)^{''}
\right) \right] \right)\;, \nonumber \\[0.3cm]
& \bar{\rho}(t,p_B^2,q^2,\szerot,\tzerot) & \;=\; &\;-\;  \frac{1}{\pi} \left[     \frac{\Ima_t \rho_2}{(s-p_B^2)(s-t)} +
\frac{1}{2} \frac{1}{s-t} \left( \frac{\Ima_t \rho_3}{s-p_B^2} \right)^{'}\,\,\right]\Bigg|_{s = \bsup{t}}\nonumber\\[0.3cm]
&                   && \;-\;  \frac{1}{\pi} \left[\frac{1}{2}  \frac{t-q^2}{\bsup{t}-t}  \partial_t\left(\frac{\Ima_t \rho_3}{(t-q^2)(\bsup{t}-p_B^2)(1-\partial_t\bsup{t})}\right)\right]  \;,\nonumber \\[0.3cm]
& \tilde{\rho}(s,q^2) & \;=\; & \left[\frac{\Rea \rho_1}{s-q^2} -   \left(  \frac{\Rea \rho_2}{t-q^2}  \right)^{'} +
\frac{1}{2}  \left(  \frac{\Rea \rho_3}{t-q^2}  \right)^{''} \right]_{t \to s}\;,   \nonumber 
\end{alignat}
\begin{alignat}{2}
& C( p_B^2,q^2,\szerot,\tzerot)     &\;=\;&-\frac{1}{\pi}\left[\frac{1}{2}\frac{\Ima_t \rho_3\big|_{s = \bsup{t}}}{(t-q^2)(\bsup{t}-p_B^2)(\bsup{t}-t)(1-\partial_t\bsup{t})}\right]\Bigg|_{t = m_b^2}^{t = \btup{m_b^2}}\nonumber\\[0.3cm]
&&&+\frac{\szerot^a}{\szerot^a+\tzerot^a}\Bigg\{\frac{\Rea\rho_2\big|_{s,t\to\Sred}}{(\Sred-p_B^2)(\Sred-q^2)}+\frac{1}{2}\partial_t\left(\frac{\Rea\rho_3}{(\Sred-p_B^2)(t-q^2)}\right)\Bigg|_{s,t\to\Sred}\nonumber\\[0.3cm]
&&&-\frac{1}{2}\frac{\szerot^a}{\szerot^a+\tzerot^a}\left(\frac{\Rea\rho_3\big|_{t\to\btup{s}}}{(s-p_B^2)(\btup{s}-q^2)}\right)'\Bigg|_{s\to\Sred}\Bigg\}\;,
\end{alignat}
where the prime denotes a derivative w.r.t. the variable $s$ and $\rho_i \equiv \rho_i(s,t)$. Above we have utilised the fact that $\Ima_t \rho_i(m_b^2,t)=\left(\Ima_t \rho_i(s,t)\right)'|_{s\to m_b^2}=0$. Application of the principal part to the double integral of \eqref{eq:double_disp_2} leads to a technical splitting of the integration region, which can be most clearly seen in \FIG\ref{fig:regions}. 
Schematically, one has
\begin{equation}\label{eq:PrincPart}
\int_{m_b^2}^{\btup{m_b^2}} dt\; P_s \int_{m_b^2}^{\bsup{t}}ds\equiv\int_{m_b^2}^{\Sred}dt\left(\int_{m_b^2}^{t-\epsilon}ds+\int^{\bsup{t}}_{t+\epsilon}ds\right)+\int_{\Sred}^{\btup{m_b^2}}dt\int_{m_b^2}^{\bsup{t}}ds\;,
\end{equation}
which corresponds to triangles B, A, and C of \FIG\ref{fig:regions} respectively.

\section{Subtracted Borel Transformation of Tree Level DA Terms}
\label{app:double-borel}

We're faced with the problem of finding the double Borel transformation of the following 
generic function ($\ell=0,1$)
 \begin{equation}
 \label{eq:F}
  {F}_{n,\ell}(p_B^2,q^2)  \equiv  \int_0^1 du \frac{(q^2)^{\ell}f_{n}(u)}{\den^n} \;,
 \end{equation}
 with $\den \equiv m_b^2 - u p_B^2 - \bar u q^2$ and $f_n(u)$  some DA  multiplying $u$-dependent prefactors.
 We explain the meaning of the silent label $n$ further below. 
 The formal solution is straightforward 
 \begin{alignat}{2}
 &\hat {F}_{n,\ell}(M_1^2,M_2^2) &\;\equiv\;& \mathcal{B}_{sub. M_2^2}^{q^2}\mathcal{B}_{sub.M_1^2}^{p_B^2}  {F}_{n,\ell}(p_B^2,q^2) \nonumber \\[0.1cm]
&  &\;=\;&  \int_{m_b^2}^{\bsup{m_b^2}} ds \int_{m_b^2}^{\btup{s}} dt 
 e^{-(\frac{s}{M_1^2} +\frac{t}{M_2^2})  }
 \rho_{F_{n,\ell}}(s,t)  \;,
 \end{alignat}
where $\bsup{t}$ and $\btup{s}$ are defined in \SEC\ref{sec:duality} and  
 $(2 \pi i)^2  \rho_{F_{n,\ell}}((s,t)  =  \disc_s \disc_t {F}_{n,\ell}(s,t)$ is the density of the 
 double dispersion representation of 
  \begin{equation}
  {F}_{n,\ell}(p_B^2,q^2) =  \int_{m_b^2}^\infty ds \int_{m_b^2}^\infty dt 
  \frac{ \rho_{F_{n,\ell}}(s,t)  }{(s-p_B^2)(t-q^2)}  \;.
 \end{equation}
If one commits to a specific function $f(u)$ the $du$-integral can be done and its double dispersion integral 
can be worked out in a relatively straightforward manner. 
In the literature the case $F^{(0)}_1$ has been worked out more generically \cite{Belyaev:1994zk}
which we generalise to $F^{(0,1)}_n$.
The function  $f_n$ is expanded, anticipating a change of variable, as
\begin{equation}
\label{eq:f}
f_n(u) =  \sum_{k \geq 0} c_{\tilde{k}} \bar{u}^{\tilde{k}}  \;, \quad \tk \equiv k + (n-1) \;,
\end{equation}
and
\begin{equation}
f^{(n-1)}_n(u) =  \left( \frac{d}{d u} \right)^{n-1} f_n(u) = 
(-1)^{(n-1)}    \sum_{k \geq 0} \bar{c}_{k} \bar{u}^{k}  \;, 
\end{equation}
where  $\bar c_k \equiv    \frac{\tk !}{k!} c_{\tk}$.
Above we have assumed that $f_n(u) \propto (u\bar u)^{n-1} ( 1 + \ORD(u,\bar u))$ which is a sufficient condition 
for the function  ${F}_{n,\ell}(p_B^2,q^2)$  \eqref{eq:F} to be free from 
$1/(p_B^2 -q^2)$ singularities.\footnote{There are some cases where this condition is not met do to the presence of $\ln u$ and $\ln \bar u$ terms, namely $\{ \mathbb A , {\cal T}^{(1)}_{1,3} \}$ and the mass corrections to $\{\tilde{\Psi}_{(a)}, \Psi^{(1)}_{(v)}\}$, for which an accurate polynomial fit can be made.}
The first dispersion representation can be obtained by a change of variable
\begin{equation}
u = \frac{m_b^2 - q^2}{s-q^2} \;, \quad \bar u = \frac{s- m_b^2 }{s-q^2} \;,
\end{equation}
 for which 
 \begin{equation}
  {F}_{n,0}(p_B^2,q^2)  =  \frac{1}{\Ga[n]} \sum_{k \geq 0} \bar c_k  \int_{m_b^2}^\infty ds
  \frac{  (s-m_b^2)^k}{(s-p_B^2)(s-q^2)^{1+\tk}} \;.
 \end{equation}
 At this level any further singularities are induced by $1/(s-q^2)^{1+\tk}$ and, 
 as discussed in the previous section,  correspond to so-called
 second type singularities. These singularities cannot appear for ${F}_{n,0}(p_B^2,q^2)$
 itself which is a fact that we have used in making the specific ansatz \eqref{eq:f}.
 The double dispersion relation then reads
 \begin{equation}
  {F}_{n,0}(p_B^2,q^2)  =  \frac{1}{\Ga[n]} \sum_{k \geq 0} \frac{\bar c_k(-1)^{\tk}}{\tk!}  \int_{m_b^2}^\infty ds
   \int_{m_b^2}^\infty dt
  \frac{  (s-m_b^2)^k \de^{(\tk)}(s-t) }{(s-p_B^2)(t-q^2)} \;,
 \end{equation}
 and its Borel subtracted form assumes the form  
 \begin{equation}
 \hat {F}_{n,0}(M_1^2,M_2^2) = 
  \frac{1}{\Ga[n]} \sum_{k \geq 0} \frac{\bar c_k(-1)^{\tk}}{\tk!} 
  \int_{m_b^2}^{ \bsup{m_b^2}} ds \int_{m_b^2}^{ \btup{s} } dt 
 \,e^{-\left(\!\frac{s}{M_1^2} +\frac{t}{M_2^2}\!\right)  }  (s-m_b^2)^k \de^{(\tk)}(s-t)  \;.
 \end{equation}
 We further decompose 
 \begin{equation}
  \hat {F}_{n,0} =   I[ \hat {F}]_{n,0}  +  \de[ \hat {F}]_{n,0} \;,
 \end{equation}
 where $I[\dots]$ and $\de[\dots]$ correspond to the integral and boundary terms that arise from 
 integration by parts. The former are easily evaluated to 
  \begin{alignat}{2}
  \label{eq:I} 
I[\hat{F}]_{n,0}  =\;&  \frac{(\hat{M}^2)^{2-n} e^{-\hat{m}^2_b }}{\Gamma[n] }   
\sum_{k \geq 0}  c_{\tilde{k}}\, \bar u_0^{\tilde{k}}\,  \left(1 - \Omega_{k+1,\Sred}\right) \;, \\[0.1cm]
I[\hat{F}]_{n,1} =\;&   \frac{(\hat{M}^2)^{3-n} e^{-\hat{m}^2_b}} {\Gamma[n]} \sum_{k \geq 0}  c_{\tilde{k}} \, 
\bar{u}_0^{\tilde{k}}
\left[  \left( \hat{m}^2_b    - \tilde{k} \bar{u}_0^{-1} \right)   \left(1 - \Omega_{k+1,\Sred}\right)  + 
(k+1  )   \left(1- \Omega_{k+2,\Sredhat}  \right)\right] \;, \nonumber
 \end{alignat}
where
 \begin{equation}
 \label{eq:u0Mhat}
u_0=\frac{M_2^2}{M_1^2+M_2^2} \;, \qquad \hat{M}^2=\frac{M_2^2 M_1^2}{M_1^2+M_2^2} \;,
\end{equation} 
 and
 \begin{equation}
 \Omega_{k,\Sredhat} = \frac{\Gamma[k,{ \Sredhat  -\hat{m}_b^2}]}{\Gamma[k]} \;,
 \end{equation}
with $\Sred$ defined in \eqref{eq:t0} and $\hat{m}_b^2 \equiv m_b^2/\hat{M}^2$. Above we have given the result for $\hat{F}_{n,1}$ in addition which does not pose any new technical challenges as one can simply replace $q^2 = s - (s-q^2)$ and treat the two terms separately.
 
 The boundary terms evaluate to 
 \begin{align}
 \label{eq:de}
\de[\hat{F}]_{n,0}  =&  \frac{1}{\Gamma(n)} \sum_{k \geq 0} \bar{c}_k  X_{\tk k}[1]  \;, 
 \nonumber \\[0.1cm]
\de[\hat{F}]_{n,1} =&  \frac{1}{\Gamma(n)} \sum_{k \geq 0}\bar{c}_k \left(    X_{\tk k}[s]- X_{(\tk-1) k}[1] \right) \;,
 \end{align} 
  $X_{\tk k}[ g(s) ]$ is the functional 
 \begin{equation}
 \label{eq:Xkk}
X_{\tk k}[ g(s) ] = \sum_{l  = 1}^{\tilde{k}} \frac{(M_2^2)^{1-l}}{\tilde{k}!}\left(\frac{\szerot^{a}}{\szerot^{a}+\tzerot^{a}}\right)^{\tilde{k}-l+1}
\partial_s^{\tilde{k}-l} \!\left[ e^{-\left( \frac{s}{M_1^2} + \frac{\btup{s}}{M_2^2} \right)}  (s-m_b^2)^k g(s)  \right]_{s= \Sred} \;.
 \end{equation}

 For further comparison with the literature 
we adopt the $\szerot,\tzerot \to \infty$ limit, for which $\Omega_{k,\Sredhat} \to 0$, 
to find
\begin{align}
\hat{F}_{n,0} \xrightarrow{\szerot,\tzerot \to \infty }\;&  \frac{(\hat{M}^2)^{2-n}e^{-\hat{m}^2_b }}{\Gamma[n]} f(  u_0)    \;, \nonumber \\[0.1cm]
\hat{F}_{n,1} \xrightarrow{\szerot,\tzerot \to \infty }\;&  \frac{(\hat{M}^2)^{3-n}e^{-\hat{m}^2_b }}{\Gamma[n]}  \left(f( u_0)(\hat{m}^2_b \pl 2\mi n ) \pl 
(1-\bar u_0)    f'( u_0 )     \right)\;,  
 \end{align}
where we used $f( u_0) =  \sum_{k \geq 0} c_{\tilde{k}} \bar{u}_0^{\tilde{k}}  $ and 
 $f'( u_0) = - \sum_{k \geq 0} c_{\tilde{k}} \tilde{k} \bar{u}_0^{\tilde{k}-1}$. 
 
 \subsection{The special case \texorpdfstring{$a=1$}{} and \texorpdfstring{$\szerot = \tzerot$}{}, \texorpdfstring{$M_1^2 = M_2^2$}{} }
 \label{app:special}
 
 For the case $a = 1$ and $\szerot = \tzerot$, $M_1^2 = M_2^2 \equiv 2  \Mbar^2 $ with 
 $\hat{M}^2 \to \Mbar^2$ and $u_0 \to 1/2$,  which is the one considered in the literature  \cite{Belyaev:1994zk}, there are miraculous simplifications. First the exponential factor in \eqref{eq:Xkk} becomes 
 $s$-independent  and \eqref{eq:de} assumes a more manageable form,
  \begin{alignat}{2}
 \label{eq:de2}
& \de[\hat{F}]_{n,0}  &\;\xrightarrow{a=1,M_1^2 = M_2^2 } \;&  
  \frac{(\hat{M}^2)^{2-n} e^{-\hat{m}^2_b }}{\Gamma[n] } 
\sum_{k \geq 0}   \frac{c_{\tilde{k}}}{2^{\tilde{k}}}\,  (\Omega_{k+1,\hat{s}_0} -\delta_{n1}\,\Omega_{1,\hat{s}_0} )\;, 
 \nonumber \\[0.1cm]
& \de[\hat{F}]_{n,1} &\;\xrightarrow{a=1,M_1^2 = M_2^2 } \;&    \frac{(\hat{M}^2)^{3-n} e^{-\hat{m}^2_b}} {\Gamma[n]} \sum_{k \geq 0}   \frac{c_{\tilde{k}}}{2^{\tilde{k}}}\, \left(  (k \pl 1) ( \Omega_{k+2,\hat{s}_0}  - \delta_{n2}\Omega_{1 ,\hat{s}_0}- \delta_{n1}\Omega_{2 ,\hat{s}_0} )  + \right.  \nonumber  \\[0.1cm]
 & &  & \left. \hat{m}^2 ( \Omega_{k+1,\hat{s}_0}  - \delta_{n1}\Omega_{1 ,\hat{s}_0} )  -2  \tilde{k}    
 ( \Omega_{k+1,\hat{s}_0}  - \delta_{n2}\Omega_{1 ,\hat{s}_0}- \delta_{n1}\Omega_{2 ,\hat{s}_0} )    \right)  \;,
 \end{alignat} 
where $\hat{s}_0=\szerot/2\hat{M}^2$.
Secondly, by  adding \eqref{eq:I} and \eqref{eq:de2} we arrive at a form where 
 \begin{alignat}{2}
\!\!   \hat{F}_{n,0}  =\;&  \frac{(\hat{M}^2)^{2-n} e^{-\hat{m}^2_b }}{\Gamma[n] } 
\sum_{k \geq 0}  \frac{c_{\tilde{k}}}{2^{\tilde{k}}} \,   (1 - \delta_{n1}\,\Omega_{1,\hat{s}_0}) = 
 \frac{(\hat{M}^2)^{2-n} }{(n\mi1)! } f_n\!\left(\frac{1}{2}\right) (e^{-\hat{m}^2_b } -  \delta_{n1} e^{-\hat{s}_0 })  \nonumber  
  \;,    \\[0.2cm]
\!\!  \hat{F}_{n,1} =\;& \frac{(\hat{M}^2)^{3-n} e^{-\hat{m}^2_b}} {\Gamma[n]} \sum_{k \geq 0}  \frac{c_{\tk}}{2^{\tilde{k}}}\left(   \hat{m}^2_b ( 1\mi \delta_{n1}\Omega_{1 ,\hat{s}_0} )  +   (2-n - \tk )   (1  \mi \delta_{n2}\Omega_{1 ,\hat{s}_0} \mi  \delta_{n1}\Omega_{2 ,\hat{s}_0}) \right)   \nonumber  \\[0.1cm]
  = \;&  \frac{(\hat{M}^2)^{3-n} }{(n\mi1)! }\Bigg[  \hat{m}_b^2 f_n\!\left(\frac{1}{2}\right)  (e^{-\hat{m}^2_b } \mi   \delta_{n1} e^{-\hat{s}_0 }) \nonumber\\[0.1cm]
&\qquad+  \left( (2\mi n)  f_n\!\left(\frac{1}{2}\right) + \frac{f_n'(\frac{1}{2})}{2} \right)\left( e^{-\hat{m}^2_b } \mi e^{-\hat{s}_0 } (   \delta_{n2}  \pl
 \delta_{n1} (1 \pl \hat{s}_0 \mi  \hat{m}_b^2)    \right) 
 \Bigg]  \;,
 \end{alignat}
 for which  the $k$-dependence in the $\Omega$-terms cancels!
 It is remarkable that for this special case the continuum subtraction vanishes for $n>1$ $(n>2)$ in $\hat{F}_{n,0}$ $(\hat{F}_{n,1})$
 and accidentally renders some results in the literature, where continuum subtractions 
 have been neglected, more accurate. 
 Note that $\hat{F}_{1,0}$  has previously been computed in \APP B of \cite{Belyaev:1994zk}
 and we agree with their result.  

\bibliographystyle{utphys}
\bibliography{../Refs-dropbox/References_FF-Bgamma.bib}

\end{document}